\documentclass{aastex631}

\newcommand{\B}[1]{{\color{blue}{#1}}}
\definecolor{ForrestGreen}{rgb}{0.133,0.545,0.133}
\newcommand{\G}[1]{{\color{ForrestGreen}#1}}

\usepackage{CJK}

\graphicspath{{./}{figures/}}

\begin{document}

\title{Magnetic Dip Found in a Quiescent Prominence Foot via Observation and Simulation}

\begin{CJK*}{UTF8}{gbsn}
\correspondingauthor{Huadong Chen}
\email{hdchen@nao.cas.cn}
\correspondingauthor{Chun Xia}
\email{chun.xia@ynu.edu.cn}

%\author{Chen et al.}
\author[0000-0001-6076-9370]{Huadong Chen (陈华东)}
\affiliation{State Key Laboratory of Solar Activity and Space Weather, 
     National Astronomical Observatories, 
     Chinese Academy of Sciences, 
      Beijing 100101, People's Republic of China}
\affiliation{State Key Laboratory of Solar Activity and Space Weather, 
     National Space Science Center, 
     Chinese Academy of Sciences, 
      Beijing 100190, People's Republic of China}
\affiliation{University of Chinese Academy of Sciences, Beijing 100049, People's Republic of China}

\author[0000-0002-7153-4304]{Chun Xia (夏莼)}
\affiliation{School of Physics and Astronomy, Yunnan University, Kunming 650500, People's Republic of China}

\author[0000-0002-5431-6065]{Suli Ma (马素丽)}
\affiliation{State Key Laboratory of Solar Activity and Space Weather, National Space Science Center, Chinese Academy of Sciences, Beijing 100190, People's Republic of China}

\author[0000-0001-9647-2149]{Yingna Su (宿英娜)}
\affiliation{Key Laboratory of DMSA, Purple Mountain Observatory, Chinese Academy of Sciences, Nanjing 210008, People's Republic of China}

%\author{Hui Tian}
%\affiliation{School of Earth and Space Sciences, Peking University, Beijing 100871, People's Republic of China}

\author[0000-0001-8228-565X]{Guiping Zhou (周桂萍)}
\affiliation{State Key Laboratory of Solar Activity and Space Weather, 
     National Astronomical Observatories, 
     Chinese Academy of Sciences, 
      Beijing 100101, People's Republic of China}
\affiliation{University of Chinese Academy of Sciences, Beijing 100049, People's Republic of China}

\author[0000-0003-3621-6690]{Eric Priest}
\affiliation{School of Mathematics and Statistics, University of St Andrews, St Andrews, Fife KY16 9SS, UK}

\author[0000-0001-9315-7899]{Lyndsay Fletcher}
\affiliation{SUPA School of Physics and Astronomy, University of Glasgow, Glasgow, G12 8QQ, UK}
\affiliation{Rosseland Centre for Solar Physics University of Oslo, P.O.Box 1029 Blindern, NO-0315 Oslo,  Norway}

\author[0000-0001-9493-4418]{Yuandeng Shen (申远灯)}
\affiliation{State Key Laboratory of Solar Activity and Space Weather, School of Aerospace, Harbin Institute of Technology,, Shenzhen 518055, People's Republic of China}
\affiliation{Shenzhen Key Laboratory of Numerical Prediction for Space Storm, Harbin Institute of Technology, Shenzhen 518055, People's Republic of China}

\author{Weining Tu (屠炜宁)}
\affiliation{Astronomy Department, Beijing Normal University, Beijing 100875, People's Republic of China}

%\author[0000-0002-5740-8803]{Bernhard Kliem}
%\author{Bernhard Kliem}
%\affiliation{Institute of Physics and Astronomy, University of Potsdam, Potsdam 14476, Germany}
\author{Wei Wang (王威)}
\affiliation{State Key Laboratory of Solar Activity and Space Weather, National Space Science Center, Chinese Academy of Sciences, Beijing 100190, People's Republic of China}
\affiliation{University of Chinese Academy of Sciences, Beijing 100049, People's Republic of China}

\author{Jun Zhang (张军)}
%\author{Jun Zhang}
\affiliation{School of Physics and Optoelectronics Engineering, Anhui University, Hefei 230601, People's Republic of China}
%\affiliation{National Astronomical Observatories, 
 %    Chinese Academy of Sciences, 
  %    Beijing 100101, People's Republic of China}

%-------------------------------------------------------
%\nocollaboration{1}
%\collaboration{1}{(LaTeX collaboration)}



\begin{abstract}
Solar prominences (or filaments) are cooler and denser plasma suspended in the much hotter and rarefied solar corona.
When viewed on the solar disc filament barbs or feet protrude laterally from filament spine.
When observed at the limb of the Sun, they reach into the chromosphere or even further down.
%Their corresponding prominence feet at the solar limb are observed to extend down to the photosphere.
%lateral feet of prominences are observed to extend down to the lower solar atmosphere.
For a long time, the magnetic field orientation of barbs has remained a mystery due to the paradox that the barbs possess vertical fine structures and flows but are likely to be supported in a horizontal magnetic field.
%often have horizontal fields suggested by the observations.
%of the horizontal field of barbs suggested by the observations and the usual appearance of vertical fine structure and flows in the barbs at the limb.
Here we present unambiguous observations of a magnetic dip in a quiescent prominence foot with an upward-curved field.
%highly suggestive
%show an unambiguous magnetic dip in a quiescent prominence foot. 
That is indicated by the horizontal bidirectional outflows probably produced by magnetic reconnection between the fields of a tiny erupting filament and those in a prominence foot. 
The altitude at the bottom of the dip is $\sim$30 Mm. At the edge of the prominence foot, the angle between the dip field and the local horizontal is $\sim$4\degr. 
Additionally, the curvature radius of the dip bottom is estimated to be around 73 Mm.
We also conduct magnetofrictional simulation to self-consistently form
a large-scale magnetic flux rope with magnetic dips resembling the spine and feet of the quiescent prominence. 
%or near
%when a small erupting filament impacted on the prominence foot.
The observations shed light on the field structure of prominences which is crucial for the instability that accounts for the eruption of prominences and coronal mass ejections. 
%Solar prominences are cooler and denser plasma suspended in the much hotter and rarefied solar corona. Most of them look like long dark “filament” due to absorption of the background radiation when they are projected on the solar disk. Filament barbs protrude laterally from the filament spine and when observed closer to the limb, some of these lateral feet of prominences are observed to extend from the spine down and end on arcs, which are thought to represent the presence of magnetic bubbles at the prominence base. For a long time the orientation of the magnetic field in the barbs remains a mystery due to the paradox of the horizontal field of barbs suggested by some observations and the appearance of vertical fine structure and flows in the barbs at the limb. Here we show an unambiguous magnetic dip structure in a quiescent prominence foot. The upward-curved field structure is indicated by the simultaneous bidirectional outflows in the aftermath of magnetic reconnection that occurred when a small erupting filament impacted on the prominence foot. The observations shed light on the structure of the prominence magnetic field and have implications for the instability that accounts for the eruption of prominences and coronal mass ejections.
\end{abstract}


\section{Introduction} \label{sec:intro}
The magnetic structure of solar prominences or filaments is a key issue in solar physics \citep[][]{tandberg95}. 
It is important not only for understanding how the prominence mass is supported in the corona, but also for clarifying the instability mechanisms of filament eruptions and the effects of eruptive filaments on space weather.
Direct field measurements of prominences are rare and arduous to make and interpret \citep[][]{vial15}.
%(??Chapter 8 in the book of solar prominence).
They have shown that quiescent prominences mostly possess horizontal magnetic fields with a field strength of 10--20 G on average and an inclination angle of $\sim$40$\degr$ to the long axis of the prominence \citep[e.g.,][]{bommier98, schmieder15, mackay20}.
%The field strength of quiescent prominences is usually 3--15 G.
%give some values and approximate orientation, however, whether the dip structure indeed exist in the barbs has not been conclusive.

Regarding the magnetic nature of filaments, the magnetic structure of filament barbs protruding laterally from the spine towards the solar surface is most controversial \citep[e.g.,][]{aulanier98b, xia14a}. Except for the dynamic short-lived barbs \citep{ouyang20}, long-lived filament barbs observed on the solar disk and prominence feet observed on the solar limb are thought to have one-to-one correspondence \citep{suy12}, so the foot and barb actually refer to the same prominence structure, which is the case in this study, and we use these two terminologies interchangeably in this work.
%protruding laterally from the spine
%Filament barbs protrude laterally from the filament spine like ramps off an elevated highway.  
It has been observed that the end points of filament barbs are close to small parasitic magnetic fields on each side of filaments \citep[][]{martin98} or magnetic network at boundaries of supergranules  \citep[][]{zhou21}, but their detailed association is not yet fully clear.
%demonstrated that the barbs corresponding to prominence feet are associated with small parasitic magnetic fields with minority polarities being opposite to the majority polarity of the network magnetic fields on each side of a filament$^8$. 
Vertical fine structure and vertical flows are often observed in prominence feet at the limb.
Assuming that the flows of filament plasma are guided by magnetic field, some studies suggest that the field lines in barbs are predominantly vertical and directly connect filament spines to the photosphere \citep[][]{zirker98}, while other observations indicate that prominence feet may harbor helical magnetic fields connecting the prominence spine downward to the solar surface \citep[][]{suy12, martinez15}.

%are field-aligned in a frozen-in condition, 
%the filament spine is connected with the photosphere through vertical fields in the barbs???. 

However, the observed velocities of the up- or down-flows in filament barbs are about 10 km s$^{-1}$, much less than the free-fall speeds corresponding to the filament heights \citep[e.g.,][]{zirker98, chae10, shen15}, which would be expected for vertical flows along magnetic field lines.
H${\alpha}$ Doppler shift observations from Meudon observatory indicate that the velocity vectors of the apparent vertical flows in a quiescent prominence have substantial horizontal components \citep[][]{schmieder10}.
%consistent with flows tied to largely horizontal magnetic fields (Schmieder et al. 2010). 
On the other hand, the vertical extent of a barb is much larger than the gravitational scale height of prominence plasma ($\sim$200 km), also implying that the prominence plasma must somehow be supported against gravity.
Some observational studies have presented the close association of prominences and coronal cavities and provided possible evidence of concave magnetic field structures -- magnetic dips as the U-shape horns at the bottom of cavities above prominence feet \citep[e.g.,][]{regnier11, berger12}.

Using the idea of magnetic dips, some linear or nonlinear force-free flux rope models \citep[e.g.,][]{aulanier98a, aulanier98b, van04, suyn12} have successfully produced three-dimensional configurations of filaments that can naturally explain the morphology of both filament spines and barbs.
According to the models, barbs represent cool matter residing in small dips, which are caused by the local distortion of a large-scale flux rope due to the introduction of parasitic polarities onto a large-scale bipolar photospheric flux distribution. This produces secondary photospheric polarity inversion lines around the parasitic magnetic elements, and is supported by results from the comparison of Big Bear Solar Observatory (BBSO) H${\alpha}$ images with magnetograms taken by SOHO MDI \citep[][]{chae05}.
\citet{regnier04} investigated magnetic dips in a small active region filament from extrapolated nonlinear force free magnetic field and found the distinction between dips in twisted flux tubes and quadrupolar configurations. At a larger scale, the ``quadrupolar'' dips may resemble those in quiescent filament barbs.

%obtained from the non-constant-αforce-free field hypothesis using a photospheric vector magnetogram
%study of magntic dips in a filament with the distinction between dips in twisted flux tubes and "quadrupolar" dips.

%Solar and Heliospheric Observatory (SOHO) Michelson Doppler Imager (MDI) 

The evidence from theoretical models and observations suggests that the vertical structure of prominence feet or filament barbs may indicate a buildup of dips in more or less horizontal magnetic field \citep[e.g.,][]{regnier04, lopez06, dudik12, lil13, suyn15, wang16, gunar18, guo21, barczynski21, yangb24}. However, as of now, there have been no clear observations of magnetic dip structures in prominence feet \citep[e.g.,][]{chae10, labrosse10, mackay10, parenti14, gibson18, chenp20}. 
In this study, we remedy this gap by analyzing a filament eruption event on April 8, 2014. 
During this event, a small erupting filament appeared to reconnect with the overlying fields of a large-scale quiescent prominence's foot. The reconnection produced bright bidirectional outflows moving along upwardly curving paths with respect to the local horizontal line in UV and EUV intensity images, strongly suggesting the presence of a magnetic dip structure across the prominence foot.

%initially interacted with a magnetic arcade under a large-scale prominence foot and then 

%Many results from theoretical models and observations tend to support the idea that the vertical structure of prominence feet or filament barbs may represent a pile-up of dips in more or less horizontal magnetic field lines 
% in a three-dimensional perspective. 
%The second interaction produced bright bidirectional reconnection outflows in the UV and EUV intensity images. They moved along gradually oblique upward directions with respect to the local horizontal line, which present an unambiguous magnetic dip structure across the prominence foot.

%present an unambiguous 
%slightly oblique upward directions

\section{Observational Data and Numerical Methods} \label{sec:obser}
The Atmospheric Imaging Assembly \citep[AIA;][]{lemen12} on board Solar Dynamics Observatory \citep[SDO;][]{pesnell12}, provides full-disk intensity images up to 0.5 R$_{\sun}$ above the solar limb with 0\farcs6 pixel size and 12s cadence in seven EUV channels. 
We used the AIA 193 \AA\ intensity data with a cadence of 60~s for the analyses of horizontal oscillations and vertical flows in the large-scale filament's barb.
One longitudinal magnetogram on April 13 with a 0\farcs5 plate scale from the Helioseismic and Magnetic Imager \citep[HMI;][]{scherrer12, schou12} on board SDO was utilized to show the photospheric magnetic field below the large-scale filament.
The H${\alpha}$ line-center intensity data were supplied by the Global Oscillation Network Group \citep[GONG;][]{harvey96} and Kanzelh\"{o}he Solar Observatory (KSO) with a pixel size of $\sim$1\farcs0 and a cadence of 60 and 20~s, respectively.
The Extreme Ultraviolet Imager on board the Solar Terrestrial Relations Observatory \citep[EUVI/STEREO;][]{wuelser04, kaiser08} provided us the 304 \AA\ intensity data with a spatial scale of 1\farcs6 and a cadence of $\sim$10~minutes from behind the solar disk.

For this event, the Interface Region Imaging Spectrometer \citep[IRIS;][]{depontieu14} slit-jaw imager (SJI) supplied the 1400 \AA\ and 2796 \AA\ intensity images with a spatial scale of 0\farcs33 and a cadence of 18 s. The IRIS spectral data are taken in a sit-and-stare raster mode with 9 second cadence and a spectral resolution of $\sim$0.025 \AA.
Because the event occurred at the solar limb, the usual method using the spectral lines such as \ion{O}{1} 1355.5977 \AA\ and \ion{Fe}{2} 1392.817 \AA\ \citep[e.g.,][]{tian15}, is not applicable for the absolute wavelength calibration.
The difference between the centroid of the total 1394 \AA\ line profile averaged over the whole slit and 1393.76 \AA\ is utilized for the absolute wavelength calibration of the \ion{Si}{4} 1394 \AA\ line spectra. A slight blueshift of $\sim$2 km s$^{-1}$, which might be caused by the solar rotation, is found from this wavelength calibration method.
After the absolute wavelength calibration, we applied a single-Gaussian fit to the spectral data and obtained the temporal evolutions of the peak intensity, Doppler shift, and line width of the \ion{Si}{4} 1394 \AA\ spectral line.

Our numerical simulation box is a spherical coordinates domain in the ranges of $1R_\odot<r<1.7R_\odot$, $97.2^\circ<\theta<144^\circ$, $0^\circ<\phi<72^\circ$, discretized by 5-layer adaptive mesh refinement (AMR) grids with a logarithmic stretch in the radial direction and 
an effective resolution of $1024\times512\times512$ cells. The limits of the domain cover the ranges of the observed filament, and are not too large to be well resolved by the numerical grids, with the finest cell size of about 370 km, 1116 km, and 1720 km in the radial, latitudinal, and longitudinal directions near photosphere. From a bipolar photospheric magnetogram, we extrapolate potential magnetic fields using the PDFI\_SS software \citep{Fisher2020} as the initial condition. Using the MPI-AMRVAC 3.0 \citep{Xia2018,Keppens2023}, we performed a magnetofrictional simulation to solve the ideal magnetic induction equation 
$\frac{\partial \mathbf{B}}{\partial t}=\nabla\times(\mathbf{v}\times\mathbf{B})$ with magnetofrictional velocity 
$\mathbf{v}=\mathbf{J}\times\mathbf{B}/(\nu_0 B^2)$, the viscous coefficient $\nu_0 = 10^{-15}\ \text{s}\ \text{cm}^{-2}$, and 
$\mathbf{J}=\nabla\times\mathbf{B}/\mu_0$. We set periodic conditions on the longitudinal boundaries, zero-velocity conditions and 
zero-gradient extrapolated magnetic fields on the outer radial boundary and the latitudinal boundaries. On the photospheric boundary, we impose equal-gradient extrapolated magnetic fields, zero radial velocity and time-evolving supergranular horizontal velocities including rotating velocities induced by the Coriolis force \citep[see][for details]{Liu2022}. The time unit is 8.3 hr and the magnetic field unit is 2 G.

\section{The Observational Structures: Large-scale Filament, Barbs, and Small Erupting Filament} \label{sec:structures}
Figs.~1(a)--(c) show the associated large-scale quiescent filament on the solar disk in the AIA 304 \AA, 193 \AA, and GONG H${\alpha}$ waveband on April 13, 2014, which was about 5 days after the minor eruption event under investigation.
The eruption occurred at the southeast limb of the Sun. During the event, a small erupting filament (referred to as ``EF'') collided and interacted with a nearby quiescent prominence's foot. 
By examining successive AIA EUV observations from April 8 to April 13 (see Video~1) alongside GONG and KSO H${\alpha}$ data, we discovered that the prominence foot corresponds to a barb structure of the large-scale filament. 
Fig.~1 displays the filament barbs (``Barb1'' and ``Barb2'') observed on the disk on April 13, along with the corresponding prominence feet at the limb on April 8. The filament EF during this event was located near Barb1 (see Fig.~1(e)). 
In the AIA 193 \AA\ channel (Fig.~1(b)), fine dark filamentary structures can be observed near the filament barbs, which appear to correspond to vertical threads in the prominence legs at the solar limb (Fig.~1(f)). 
The two ellipses approximately indicate the locations of the bottom regions of the threads, which likely correspond to the feet (Ft1 and Ft2) of Barb1 and Barb2 on the solar disk.
A line-of-sight magnetogram from SDO HMI (Fig.~1(d)) illustrates the weak quiescent photospheric magnetic fields with different polarities located on either side of the filament channel. 
It appears that the feet Ft1 and Ft2 are positioned along the boundary of some network magnetic field.
%, suggesting that the formation of the magnetic field structures associated with the barbs may be related to the activities of the photospheric supergranulation.
Comparing panel (e) with panel (g), it is evident that the barb structures on the AIA 304 \AA\ image are significantly wider than those shown in the GONG H${\alpha}$ band. This suggests that the structures observed in the GONG H${\alpha}$ band correspond to the cooler core regions of the barbs. Additionally, the AIA 171 \AA\ intensity image, enhanced using the Multi-scale Gaussian Normalization (MGN) method \citep[][]{morgan14}, is presented in Fig. 1(f). It reveals some dip structures along the tops of the barbs and within the vertical threads of Barb1, similar to previous coronal cavity observations \citep[e.g.,][]{regnier11, berger12, suyn15, wang16, yangb24}. However, the projection effect complicates our understanding of the true spatial relationship between these dips and barbs.
In this study, we will employ two methods described below to examine the specific location of Barb1 during the EF's eruption and analyze the GONG H$\alpha$, AIA EUV and IRIS UV data to demonstrate that the EF interacted with the core of Barb1 (Section 4).

%\subsection{The Location of Barb1 at the Solar Limb on April 8} \label{sec:location}
The observations from the AIA 304 \AA\ channel clearly illustrate the process of the EF eruption. Additionally, in the AIA 171 \AA\ and 211 \AA\ wavebands, some brightenings appeared in the source region following the eruption (refer to Video~3). This helps us identify the location of the EF eruption source region.
In Fig.~1, the plus signs in panels (e) and (f) denote the location of EF eruption source, while the diamond sign in panel (g) indicates Barb1's footpoint (Ft1). It can be seen that the filament EF's source region on the disk is approximately located at (-90\fdg0$\pm$1\fdg0, -50\fdg7$\pm$0\fdg1), and the latitude of Ft1 is about -51\fdg3$\pm$0\fdg1. Based on their latitudinal relationship, the bottom of Barb1 is very close to the eruption source region. 

This event was also observed by the EUVI on board the spacecraft B of STEREO. However, due to insufficient temporal resolution, the EUVI observations do not allow for a detailed analysis of the event's evolution. By using the EUVI observations from a different perspective (see Fig. 2(a)), combined with the SDO/AIA observations, we are able to measure the positions of Barb1 and the small erupting filament EF near the solar limb. 
Fig. 2(b) and (c) illustrate the simultaneous imaging of Barb1 and EF by AIA and EUVI at 07:46 UT on April 8. We selected four points (P1-P4) on Barb1 and two points (P5 and P6) on EF for our analysis. Using the observations from the two perspectives and employing the program (scc$\_$measure.pro) from the solar software \citep[SSW;][]{freeland98}, we obtained the heliographic longitude and latitude of these six points, along with the radial distance in solar radii (see Table~1). Our measurements indicate that the average longitude of Barb1 is -92\fdg3, while the average longitude of EF is -90\fdg9, consistent with the EF source region's longitude of -90\fdg0$\pm$1\fdg0 described above.
These results show that Barb1 and EF are both close to the solar limb and the longitude difference between them is about 1\degr.
%These findings are also consistent with the results of -90\fdg4$\pm$1\fdg0 and -90\fdg0, respectively, from the calculations and analyses of the differential rotation speed. 

Additionally, since there was no significant eruption activity related to the large-scale filament between April 8 and April 13, the magnetic field structure is expected to remain stable \citep[][]{tandberg95, priest14}.
Generally, the barb structures could be evolving over time because their footpoints are located in network fields which are continuously changing as they are pushed around by granulation and supergranulation. However, it is theoretically challenging to estimate and determine the specific impact of the random motion of the photospheric granules and supergranules on the position change of the barbs footpoints.
Observationally, for the large-scale filament studied here, Barb2 is situated further west and closer to the solar disk center. The AIA EUV successive observations over these days indicate that Barb2's position within the filament has not changed significantly (see Video~1). Therefore, we hypothesize that the same stability applies to Barb1.
Assuming that Barb1 rotated rigidly along with the filament and the Sun, we can infer its longitude at the solar limb on April 8 by determining the differential rotation speed and Barb1's longitude on the solar disk on April 13.
We utilised the GONG H$\alpha$ and AIA 304 \AA\ data to establish Barb1's longitude and latitude on the solar disk around 02:20 UT on April 13 (see Fig.~3). 
It should be noted that due to the projection effect and the fact that the barb structure has a certain spatial scale, it is difficult to determine the foot of Barb1 on the solar disk very accurately. Here, we take the center of Barb1 as the reference point, indicated by the plus signs in Fig.~3.
The coordinates are (-32\fdg5$\pm$0\fdg1, -49\fdg3$\pm$0\fdg1) and (-33\fdg7$\pm$0\fdg1, -48\fdg9$\pm$0\fdg1) obtained from the H$\alpha$ and 304 \AA\ data, respectively. 
Combined with the latitude of Ft1 at the solar limb -51\fdg3, we obtain the Ft1's average latitude to be -49\fdg8$\pm$1\fdg7.
%It can be seen that the mean latitude of Barb1 at this time (-47\fdg5) differed by about 4\arcdeg\ from when it was near the limb (-51\fdg3, see Fig.~1(g)). 
%This discrepancy may primarily stem from potential inaccuracies in the measured points. Due to the projection effect, accurately pinpointing the location of Barb1's footpoint on the solar disk is quite challenging.

According to \citet{howard70}, the formula for the solar photosphere's differential rotation can be expressed as
\begin{equation}
\omega= 2.78 \times 10^{-6} - 3.51 \times 10^{-7} \sin^{2} \beta - 4.43 \times 10^{-7} \sin^{4} \beta  
 \   \  \   \  (rad  \   s^{-1}) 
 \end{equation}
 or
 \begin{equation}
 \omega= 13.762 - 1.738 \sin^{2} \beta - 2.193 \sin^{4} \beta  \   \  \   \  (deg   \   day^{-1}),
\end{equation}
where $\beta$ represents the solar latitude.
%From April 8 to April 13, the average latitude of Barb1's footpoint was approximately -49\fdg4. 
Taking $\beta$ as -49\fdg8, accordingly, the differential rotation speed is approximately 12\fdg0 $\pm$ 0\fdg14 day$^{-1}$. Based on the average longitude of Barb1 at 02:20 UT on April 13 (-33\fdg1), it can be inferred that at 07:40 UT on April 8, the longitude of Barb1 was about -90\fdg4$\pm$1\fdg0, 
which only has a difference of $\sim$2\degr\ compared to the result of -92\fdg3 obtained using the dual-perspective observation measurement method.
Since these two longitudes correspond to Barb1's foot and its overlying body, respectively, our results are quite reasonable.

%which is near the limb of the Sun and very close to the location of the small filament EF's source region \textbf{(-90\fdg0$\pm$1\fdg0, -50\fdg7$\pm$0\fdg1)}. 

The analysis above indicates that the longitude difference between Barb1 and EF is about 1\degr. At the location where EF and Barb1 collided (such as P4 in Fig.~2), this 1\degr\ difference corresponds to a distance of $\sim$8 Mm, while the GONG and KSO  H$\alpha$ data show that the width of Barb1 is $\sim$10 Mm, and the width of EF is $\sim$5 Mm in the AIA 304 \AA\ images. Considering that the magnetic field structures of Barb1 and EF may have larger spatial scales, we believe that when EF erupted at 07:40 UT on April 8, interaction occurred between Barb1 and EF and resulted in the subsequent magnetic reconnection.

%Taking into account that the spatial scales of Barb1 and EF, %longitude span of Barb1 is $\sim$5\arcdeg (see Fig.~2b), 

%\section{The Disturbance of Barb1 from the EF's Eruption in H$\alpha$} \label{sec:disturbance}
\section{Observational Evidence of Magnetic Reconnection between Barb1 and EF} \label{sec:evidences}
In Fig.~4, the GONG H$\alpha$ line-center observations from 07:40 to 08:00 UT on April 8 are displayed (also see Video~2). The small erupting filament EF was not clearly visible in the H$\alpha$ data, likely due to its small size and weak emission in H$\alpha$. 
However, when EF erupted during this period (also refer to Fig.~5 and Video~3), it seems that the prominence material in Barb1 was disturbed. Part of the material in the southern thread of Barb1 was stripped out and moved outward in a locally horizontal and oblique upward direction, which appears to be along the magnetic dip structure, as observed in the IRIS SJI 1400 \AA\ waveband (see Fig.~6). These observations provide further evidence that the core region of Barb1 interacted with the erupting filament EF.

%\subsection{EF's Eruption and Interaction with Barb1 in the AIA EUV and IRIS UV} \label{sec:interaction}
Fig.~5 and 6 show the details of the small filament eruption and its interaction with Barb1 in the AIA EUV and IRIS UV wavebands, respectively.
Due to the limited field of view (FOV) of the IRIS SJI observations, the IRIS UV images in Fig.~6(a)-(c) cover a smaller area compared to the AIA images in Fig.~5.
Before the eruption, the eruptive filament EF is very small and is hardly visible.
In addition, in both the AIA 304 \AA\ and the IRIS SJI 2796 \AA\ channels, we observe several nearly horizontal prominence structures that appear to pass in front of Barb1, as indicated by the yellow dotted lines in Fig.~5(a), (b), and Fig.~6(a). 
When the eruption began around 07:30 UT, these prominence structures were visibly pushed upward and flowed northeastward. Concurrently, some brightening areas ("Br") began to appear below these structures, indicating that magnetic reconnection might be occurring. 
As the Br was pushed aside, the erupting filament EF continued to rise and collided with the overlying prominence material and magnetic fields in Barb1, during which reconnection likely continued to occur. At an altitude of $\sim$30 Mm, we have detected simultaneous bidirectional bright outflows from the reconnection region in the AIA 304 \AA\ and 171 \AA\ lines. These reconnection outflows are referred to as two-sided-loop jets in some studies \citep[e.g.,][]{yokoyama95, zheng18, shen19, cheny24, houz24, yangj24, yangl24}. In the 171 \AA\ observations, some absorption features (indicated by the black arrow in Fig. 5(h)) in the vertical threads of Barb1 were found to be disturbed and moved behind the bright outflows. This is in good agreement with the disturbance of the prominence material caused by the EF eruption observed in the GONG H$\alpha$ band. The reconnection outflows were not visible in hotter AIA channels such as 193 \AA, 211 \AA\ and 335 \AA, suggesting that the highest temperature of the plasma in these outflows may be around 0.6 MK, which corresponds to the peak temperature response of the 171 \AA\ channel.

In Fig.~6, the IRIS SJI 1400 \AA\ data with higher spatial resolution ($\sim$0\farcs33) shows the bidirectional reconnection outflows more clearly. 
It can be seen that several groups of reconnection outflows successively took place, consistent with reconnection between the fields of filament EF and Barb1. 
The enlarged SJI 1400 \AA\ images in Fig.~6(d)-(i) show the close-ups of three examples of the reconnection outflows. 
Most importantly, we found that these reconnection outflows gradually curved upwards relative to the local horizontal orientations. Assuming the magnetic field is frozen in to the plasma, the upward-curved trajectories of the reconnection outflows strongly suggest that some magnetic dip fields existed and crossed the vertical structures in Barb1.
It is worth noting that these reconnection outflows differ from the earlier prominence material flows that moved in the northeast direction, as indicated by the yellow arrows in Fig.~5(b)-(e) and Fig.~6(b). Here are the key differences: (1) The reconnection outflows are bidirectional, whereas the earlier material flows are unidirectional; (2) Their later movement directions are clearly different--the reconnection outflows move obliquely upward, while the earlier material flows appear to move toward the solar surface; (3) The earlier material flows do not show significant signatures in the AIA 171 \AA\ and IRIS 1400 \AA\ bands. These differences suggest that the earlier flows may simply represent the movement of certain prominence materials caused by the compression of the EF eruption, whereas these outflows are consistent with materials ejected due to the reconnection of the EF with the magnetic field in prominence Barb1.

%\subsection{The Dynamics of the Erupting Filament EF, Overlying Prominence Material, and Outflows} \label{sec:dynamics}
Along the slit A--B (see Fig.~5(e)) and a curved slit C--D (see Fig.~6(g)), we produced AIA 304 \AA\ and IRIS 1400 \AA\ time-distance diagrams in Fig.~7(a) and (d), respectively.
They separately show the dynamics of the small erupting filament EF, its overlying prominence material in Barb1 and the bidirectional reconnection outflows.
It can be seen that as filament EF erupted upwards, its overlying prominence material was compressed and forced to rise by about 10 Mm and then stopped.
The temporal profiles of the projected velocity and acceleration of EF (red) and the overlying prominence material (blue) are presented in Fig.~7(b) and (c), respectively.
The filament EF underwent a rapid-acceleration phase with a maximum acceleration of 0.27$\pm$0.07 km s$^{-2}$ after the eruption onset.
At about 07:36 UT, EF reached a maximum speed of 56$\pm$3 km s$^{-1}$ and then was decelerated probably due to confinement by the overlying magnetic structures and prominence material.
After 07:48 UT, filament EF gradually faded out of the AIA 304 \AA\ images.
Six pairs of reconnection outflows (``o1''--``o6'' in Fig.~7(d)) are identified in the IRIS 1400 \AA\ time-slit map, which has been enhanced by the MGN method.
Applying linear fits to the reconnection outflow trajectories in the time-distance map, we found that the projected velocities of the outflows moving towards C are on average larger than those of the outflows towards D.
Their average values are $\sim$85 km s$^{-1}$ and $\sim$62 km s$^{-1}$, respectively.
The difference in reconnection outflow velocity along the two directions is probably due to the asymmetry in ambient outflow conditions, with that in the region of the reconnection outflows towards D being denser.
%downflow region being denser.
Another possibility is that materials travel at the same speed in both directions but along a magnetic structure which is curved in 3D, so that the projected speeds in the plane of the sky look different.

%could it also be due to material travelling at the same speed in both directions but along a magnetic structure which is curved in 3D, so that the projected speeds in the plane of the sky look different? 
 
%\subsection{Spectral Analyses of the Outflows} \label{sec:spectral}
Some reconnection outflows towards C were captured by the IRIS spectrometer slit, which is marked by the vertical black lines in the IRIS UV intensity images (Fig.~6). We mainly analyzed the spectra of the \ion{Si}{4} 1394 \AA\ line to deduce the spectral characteristics of the reconnection outflows. The 1394 \AA\ spectral line is formed at a temperature of $\sim$0.08 MK. After doing wavelength calibration, we applied a single-Gaussian fit to the spectral data and obtained the temporal evolutions of the peak intensity, Doppler shift, and line width from 07:43 to 08:10 UT, which are separately displayed in the top three panels of Fig.~8. It can be seen that the first reconnection outflow crossed the IRIS slit at about 07:45 UT. Subsequently, multiple reconnection outflows followed the first one and passed through approximately the same position of the spectrometer slit. The Doppler speeds of the reconnection outflows had significant blueshift signatures ([-5, -18] km s$^{-1}$)  before 07:55 UT and then exhibited redshifts ([7, 20] km s$^{-1}$), suggesting that different dipped field lines, some of them are oriented out of  the plane of the sky and some into the plane of the sky, participated in the reconnection with filament EF. The average Doppler speed is $\sim$11 km s$^{-1}$, much less than the projected velocities ($\sim$85 km s$^{-1}$) of the reconnection outflows, suggesting that the outflows are located close to the plane of the sky. The Doppler line widths vary in the range [5, 16] km s$^{-1}$ with a mean value of $\sim$10 km s$^{-1}$, and a mean nonthermal velocity of $\sim$6 km s$^{-1}$ probably due to turbulence or unresolved Alfv\'{e}n waves in the reconnection outflows \citep[e.g.,][]{tian14}. The IRIS SJI 1400 \AA\ intensity images of the two reconnection outflows at 07:51 and 08:08 UT and the corresponding \ion{Si}{4} 1394 \AA\ line spectra and line profiles are plotted in the bottom panels of Fig.~8. It can be seen that the 1394 \AA\ spectral line is separately blueshfited by $\sim$13 km s$^{-1}$ and redshifted by $\sim$20 km s$^{-1}$ for the two reconnection outflows.

%\subsection{Quantitative Analysis of the Magnetic Dip Structure} \label{sec:quantify}
\section{Magnetic Dip in Observations and Numerical Prominence Model} \label{sec:dip}
Figuring out the magnetic characteristics of the dip fields in the feet is helpful and important for understanding the magnetic structure and equilibrium of the prominences.
Generally, the prominence material is heated by magnetic reconnection to tens of thousands or millions of degrees and so would become fully ionized, which leads to a high electrical conductivity of the heated plasma.
%hundreds of thousands or millions of degrees 
%a temperature between
It is likely that the reconnection outflows under investigation were ejected from the reconnection region and moved along the magnetic fields because of the frozen-in condition. 
%reasonable to think
Since the Doppler speeds of the reconnection outflows are much smaller than the projected speeds, the motion trajectories of the outflows approximately reflect the real configuration of the magnetic dip fields.
% i.e., the outflows' field components towards us are far less than the others,
%, since the field component along the line of sight can be neglected. 
In Fig.~9(a), we draw the trajectory of the reconnection outflow observed by IRIS at 07:56:17 UT (see Fig.~6(g)) on the background of a simultaneous KSO H$\alpha$ intensity image.
Here, we established a cartesian system of coordinates, where the $z$-axis refers to the vertical axis of Barb1, the $y$-axis points to us, and the $x$-axis represents the horizontal direction perpendicular to Barb1.
The altitude (height) of the bottom of the dip is estimated to be $\sim$30 Mm.
The angles $\varphi$ between the dip field and the $x$-axis were calculated and plotted in Fig.~9(b).
The mean value of $\varphi$ at the edge of Barb1 is 3.9$\arcdeg$$\pm$0.4$\arcdeg$, which are denoted by the plus signs and vertical dashed lines in Fig.~9(a).
We also calculated the curvature radii projected in the plane of the sky of the dip bottom, with a mean value of 73$\pm$9 Mm.

Before reconnection, the equilibrium of the prominence material in Barb1 is approximately governed by the magnetohydrostatic force balance equation \citep[][]{kippenhahn57, priest14}
\begin{equation}
0=-\triangledown p - \rho g \hat{z} - \triangledown (\frac{B^2}  {2\mu}) + (\textbf{B} \cdot \triangledown) (\frac{\textbf{B}} {\mu}).
\end{equation}
Here, $p$, $\rho$, $g$, $\mu$ and $B$ are the gas pressure, mass density, solar gravitational acceleration, magnetic permeability and magnetic field strength, respectively.
%where $p$,  is gas pressure, $\rho$ mass density, $g$ the solar gravitational acceleration, $\mu$ magnetic permeability and $\textbf{B}$ 
%Using the vertical prominence sheet model$^{40}$, 
At the edges of Barb1, the $z$-component of Eq. (1) implies in order of magnitude that 
\begin{equation}
\rho g \approx \frac {2B_{x} B_{z}} {\mu w},
\end{equation}
%Assuming that $B_{x}$ keeps constant along the outflow
where $w$ is the width of Barb1 and $B_{x}$ is assumed uniform. 
%along the dip field.
%$B_{z}$ and is $B_{x}$ are the values at the edge of Barb1. 
Replacing $B_{z}$ with $B_{x}$$\tan\varphi$, evaluated at the edge of the barb, this implies that 
\begin{equation}
\rho g \mu w \approx 2 B_{x}^2 \tan \varphi.
\end{equation}
Given that the electron density \citep[][]{chenh20} and ionization degree \citep[][]{anzer07} of a typical prominence is 2.4 $\times$ 10$^{10}$ cm$^{-3}$ and $\sim$0.3, respectively, then $\rho$ $\approx$ 1.3 $\times$ 10$^{-10}$ kg m$^{-3}$ .
Coupled with $w$ $\approx$ 9700 km and $\varphi$ $\approx$ 4$\degr$, this implies that $B_{z} \approx 1$ G, and $B_{x} \approx 18$ G, which is a typical field strength for quiescent prominences \citep[e.g.,][]{bommier86, mackay20}. 

%\begin{equation}
%0=-\triangledown p - \rho g \hat{z} - \triangledown (\frac{B^2}  {2\mu}) + (B \cdot \triangledown) (\frac{B} {\mu}),
%\end{equation}
%where $p$ is the gas pressure, $\rho$ is mass density, $g$ is the gravitational acceleration, $B$ is magnetic field strength, and $\mu$ is permeability.
%$\rho g \approx \frac {2B_{z\infty} B_{x}} {\mu w}$.

%\subsection{Horizontal Oscillations and Vertical flows in Barb2} \label{sec:flows}
We also observed that the barbs of the large-scale filament (or prominence's feet) exhibited noticeable horizontal oscillations as well as vertical up- and down-flows, both at the solar limb and on the solar disk (see Video~5). This was particularly evident for Barb2, especially when they were on the solar disk. In Fig.~10, we present time-distance diagrams created using the intensity images at AIA 193 \AA\ from 07:00 UT to 20:00 UT on April 8 and from 04:00 UT to 17:00 UT on April 11. These diagrams were made along three slits (S1, S2, and S3) at the solar limb and two slits (S4 and S5) on the solar disk. The two sets of 193 \AA\ data have been de-rotated to 13:30 UT on April 8 and 10:30 UT on April 11, respectively. The statistical kinetics of the horizontal oscillations and vertical flows in Barb2 are summarized in Table~2.

The panels (a) and (e) in Fig.~10 display the horizontal oscillation characteristics of Barb2 at the solar limb and on the solar disk, respectively. These oscillations typically last for 2 to 3 cycles, with mean projection amplitudes, periods, and velocities measured at the solar limb being 1.4 Mm, 29 minutes, and 3.5 km s$^{-1}$, while those on the solar disk are 0.9 Mm, 30 minutes, 2.3 km s$^{-1}$. Notably, the amplitude and velocity at the solar limb are greater than those measured on the solar disk, which may be attributed to the projection effect. If we consider the amplitude and velocity values in these two directions as the vertical components of a 3D oscillation, we can estimate the true amplitude of the horizontal oscillation in the barb to be ~1.7 Mm, with a corresponding velocity of 4.2 km s$^{-1}$,  comparable to the observations of previous studies \citep[e.g.,][]{berger08, lin09, lil13}. 

Around 11:23 UT on April 11, an eruption occurred at the eastern limb of the Sun, northeast of Barb2, generating a coronal wave  \citep[e.g.,][]{liuw14, shen22, zheng24}. This wave reached Barb2 at 11:33 UT and had a significant impact, causing stronger horizontal oscillations to the vertical threads in Barb2 (indicated by the arrows in Fig.10(e)). Unlike other types of oscillations, this one was more intense, with a maximum amplitude of ~3.6 Mm, a peak velocity of ~7.4 km s$^{-1}$, and a mean period of 33 min, displaying an evolution with a gradual damping. This suggests that the physical causes of the two types of oscillations are different. Weak oscillations in Barb2 may be linked to ubiquitous disturbances to the equilibrium of the plasma within the barb's dip or the propagations of magnetoacoustic waves \citep[e.g.,][]{berger08, lin09, mackay10, lid18}, while the stronger oscillations were triggered by the coronal wave from the distant eruption  \citep[e.g.,][]{lit12, panesar13, bi14, shen14, zhangq18, luna24}.

Large-amplitude longitudinal (LAL) oscillations of material along the filament axis have been frequently observed and reported \citep[e.g.,][]{jing03, lit12, zhangq12, zhangq17}. Some observational and simulation studies have employed a pendulum model, in which gravity is the restoring force, to explain these LAL oscillations \citep[e.g.,][]{luna12a, luna12b, luna16, zhangq13, zhou18}. Given that the magnetic dip we observed in the barb may have a structure similar to the large-scale axial magnetic field of filaments, we try to investigate whether the pendulum model can also account for the horizontal oscillations of the barb. In the pendulum model, the theoretical period is determined by $P = 2\pi\sqrt{R/g}$, where R is the curvature radius of the dipped portion of the field lines and g $=$ 274 m s$^{-2}$ is the solar gravitational acceleration.
Since Barb1 and Barb2 belong to the same filament, are in close proximity to each other, and have similar spatial scales (see Fig. 1), we think that they should have similar magnetic field structures. Therefore, we simply utilize the curvature radius of the dip observed in Barb1 to estimate the theoretical oscillation period of Barb2.
The derived period is 54$\pm$3 minutes, which is nearly twice the observed oscillation period of $\sim$30 minutes. Several possible physical reasons may explain this discrepancy: (1) The oscillations in the barb may be influenced by the coupling of gravity and gas pressure. 
\citet{luna12b, luna16} indicate that in the case of a shallow dip (or for a large curvature radius), the component of gravity along the field lines is small, and the pressure force becomes important; (2) The curvature radius of the dip in the barb may vary with altitude. At higher altitudes, the curvature radius may increase. Additionally, the upward compression from the erupting filament EF could cause the dip to flatten, resulting in a larger curvature radius; (3) Barbs extend from the filament spine to the solar lower atmosphere, and magnetoacoustic waves \citep[e.g.,][]{berger08, terradas13, lid18} or mass injection \citep[e.g.,][]{yan25} from the chromosphere may also influence the oscillations in the barbs.

%According to the pendulum model

%不同的物理机制，比如重力、磁压梯度和磁张力等，来解释这种振荡。
%观测和模拟发现，在受到外在影响时，比如附近的爆发或腿足的局部加热等，

Fig. 10(b), (c), and (f) illustrate the kinetics characteristics of the vertical flows in Barb2 at the solar limb and on the solar disk, respectively. These vertical flows include both up-flows and down-flows. Due to the limitations of the code drot$\_$map.pro in SSW, which cannot de-rotate images beyond the solar limb, we tracked the position of Barb2, which shifted southeast during the timeframe of 07:00 to 20:00 UT on April 8. We created time-distance maps along the two nearly parallel slits, S1 and S2: one from 07:00 UT to 13:30 UT (shown in panel (b)) and the other from 13:30 UT to 20:00 UT (shown in panel (c)). We calculated the average projected motion distance, duration, and velocity of upflows and downflows, as presented in the right columns in Table~2. Generally, downflows exhibit longer motion distances, longer durations, and greater velocities compared to upflows. Their average motion distances, durations, and velocities are 8.9 and 5.8 Mm, 17 and 14 minutes, and 10 and 7 km s$^{-1}$, respectively. When comparing data from at the solar limb and on the solar disk, it is evident that the values of the motion distance and velocity are larger at the limb, which appears to be related to the projection effect. Moreover, the white curved lines in panel(f) indicate that upflows and downflows can sometimes occur continuously. This is clearly demonstrated in the evolution of Barb2 on the solar disk.

%\subsection{Magnetic Dip in Numerical Prominence Model}\label{sec:model}
To theoretically demonstrate the validity of magnetic dips in prominence feet, we conduct a self-consistent numerical model on the 
formation of quiescent prominence magnetic field. Starting from a bipolar potential field (Fig. 11(a)), vortical motions at supergranular 
boundaries inject positive magnetic helicity into small flux tubes, between which magnetic reconnections transfer magnetic helicity from
small flux tubes to the boundaries of larger flux tubes, thus magnetic helicity condenses at the polarity inversion line (PIL) and appear as 
strongly sheared magnetic loops (Fig. 11(b)). Due to magnetic reconnection of foot points driven by supergranular converging flows at the 
PIL, different sheared loops become helical field lines and form a magnetic flux rope (Fig. 11(c)). Magnetic reconnections occur due to 
numerical resistivity at current sheets. The magnetic flux rope grows from thin to thick and show mature structures of spine and feet at 
time 105 (Fig.~11(d)). Note that since the magnetic dip regions as gravitational potential wells can collect and hold prominence plasma, 
they are used to approximately represent prominence plasma structures. In Fig. 11(d), the green helical magnetic field line passes through 
a prominence foot at the bottom of a magnetic dip, which is the middle part of the helical field line and well resembles the observed 
magnetic dip in Fig.~6 (g). The bottom of the magnetic dip has a height of 22 Mm and the apparent curvature radius is very close to the
observed ones. Panel (e) of Fig.11 shows the top view of the same prominence and the green magnetic dip as a filament on the solar disk at
time 105. White helical field lines represent the magnetic flux rope hosting the filament. Barbs or feet protrude laterally from the spine 
to the northeast direction similar to the observed filament in Fig.~1 (c). The simulated magnetic network and the southeast--northwest 
oriented PIL roughly match the right part of the magnetogram under the observed filament (Fig.~1(d)), which is closer to the solar disk center.

\section{Discussion} \label{sec: discu}
Whether a magnetic dip exists in filament barbs or prominence feet has been a matter of debate for many years.
Some direct magnetic or velocity vector field measurements \citep[e.g.,][]{mackay20, levens16, schmieder17} of prominences suggest that horizontal magnetic components may indeed exist in prominence feet.
The vertical fine structure of prominence feet may then be an accumulation of magnetic dips, which are well reproduced in some theoretical models \citep[e.g.,][]{aulanier98a, van04, luna17}.
Observationally, however, up to now, any conspicuous dip configurations across prominence feet have not yet been unambiguously detected and presented.
%Using a flux-rope insertion method, some field extrapolations about intermediate filaments seem to support a positive answer of this question$^{17,42}$.
In this study, using high-resolution imaging and spectroscopy data from IRIS and SDO, we clearly demonstrate a magnetic dip structure in a quiescent prominence foot, which is revealed by upwardly-curved trajectories of simultaneous bidirectional outflows (or jets) driven by the reconnection between the fields of a small erupting filament and those of a prominence foot. 
The results provide highly suggestive observational evidence for dip fields in prominence feet or filament barbs.
It should be noted that the two-sided-loop jets presented by \citet{shen19} also indicate the presence of magnetic dip structures beneath a quiescent large-scale filament. 
However, the connection between these dips and the filament barbs is not clearly demonstrated or explained.
%have found highly suggestive evidence for 
%clearly demonstrate;  or close to; or close to
%On the other hand, there may also be vertical fields extending down to the photosphere in prominence foot.
%we cannot still exclude that there are no 
%In the AIA 304 \AA\ and IRIS UV imaging data, a small arcade Ar appeared at the bottom of Barb1, implying that some fields may come out from the ends of Ar and pass upwards through Barb1.
%The appearance of a small arcade Ar at the bottom of Barb1 seem to indicate some fields coming out from the ends of Ar and passing upwards through Barb1.

As for the dip fields in the foot, when and how they formed and what their relationship is to the prominence fields, are still not clear.
Some theoretical studies have modeled filament barbs forming from parasitic polarities interacting with the filament spine \citep[e.g.,][]{aulanier98b, van04}.
In our simulation, driven by the photospheric supergranular flows, a large-scale twisted magnetic flux rope forms and develops spine and feet 
structures represented by magnetic dips, with high similarity to the observed filament. This self-consistent numerical model corroborates the
magnetic dip found in the observation.
Moreover, the reconnection between the approaching opposite polarity legs of different sheared magnetic arcades along a filament channel may also produce such dipped fields \citep[e.g.,][]{van89, martens01, xia14b, chenh14, chenh15, chenh16, yan15}.
%seems to be capable of 
When cool and dense plasma accumulates in such dipped fields by injection or condensation, a filament barb or prominence foot may form.
%In this scenario, a magnetic arcade naturally appears below the dip after the reconnection, consistent with the observations presented here.
In Fig. 12(a) and (b), we have plotted the possible magnetic field structure on the lower part of a large-scale filament.
The flux rope has two groups of dipped fields, which correspond to the two barbs Barb1 and Barb2, respectively.
Some fields (dashed lines in the diagram) may be anchored in the chromosphere or photosphere around the barbs.
Fig. 12(c)--(e) show the eruption of the small filament EF and its reconnections with the overlying dip fields in Barb1.
The heated plasma was rapidly expelled from the reconnection region and flowed along the dip fields in Barb1 to form the bright reconnection outflows, as observed by IRIS and SDO/AIA.

Similar to many observations \citep[e.g.,][]{chae10, berger11, lil13, shen15, bi20, rees24, yangb24}, our observations also indicate the presence of restless upflows and downflows in the prominence feet.
Theoretical models \citep[][]{aulanier98a, aulanier98b, van04, suyn12} and the observations presented in this paper, suggest that nearly horizontal magnetic dips exist in the prominence feet or filament barbs, providing support for the dense, heavy prominence material. However, static magnetic dip structures alone are insufficient to explain the observed vertical flows.
One theoretical hypothesis posits that prominence dynamics may develop such extreme physical conditions that the usual frozen-in condition breaks down spontaneously \citep{low12a}. 
In the ``droplet'' model proposed by \citet{haerendel11}, plasma packets can squeeze themselves through the predominantly horizontal magnetic field under the influence of gravity, during which  Alfv\'{e}n waves are generated in the horizontal magnetic field and help to keep the velocity of the falling plasma constant.
Additional studies indicate that, despite the high magnetic Reynolds number within prominences, magnetic reconnection can readily occur. This happens through the spontaneous formation and dissipation of electric current sheets, which facilitate the transfer of material between prominence dips \citep[][]{petrie05, chae10, low12b}.
Moreover, some numerical simulations have examined the role of the Magnetic Rayleigh--Taylor Instability in generating upflows and downflows in prominences \citep[e.g.,][]{hillier12a, hillier12b, xia16}.
Specifically, \citet{rees24} conducted a 2.5D simulation of knot formation under the effects of Magnetic Rayleigh--Taylor Instability to explain their observations.
To fully understand the dynamic phenomena occurring within prominences, future research will require higher spatial resolution multi-waveband observations, more precise magnetic field measurements, and more MHD simulations for solar prominences.

\begin{acknowledgments}
This work is supported by the National Key R\&D Program of China 2021YFA1600502, the Strategic Priority Research Program of the Chinese Academy of Sciences (grant No. XDB0560000), and the Specialized Research Fund for State Key Laboratories of Solar Activity and Space Weather.
C.X. acknowledges the NSFC support (11803031, 12073022) and the Basic Research Program of Yunnan Province (2019FB140, 202001AW070011).
L.F. acknowledges support from grants ST/L000741/1 and ST/ X000990/1 made by UK Research and Innovation's Science and Technology Facilities Council (UKRI/STFC). 
Y.D.S. acknowledges Shenzhen Key Laboratory Launching Project (No. ZDSYS20210702140800001). 
H.D.C. was also supported by the Chinese Academy of Sciences (CAS) Scholarship.
IRIS is a NASA small explorer mission developed and operated by LMSAL with mission operations executed at NASA Ames Research center and major contributions to downlink communications funded by ESA and the Norwegian Space Centre. The SDO data are courtesy of NASA, the SDO/AIA, and SDO/HMI science teams.
The H-alpha data were acquired by GONG instruments operated by NISP/NSO/AURA/NSF with contribution from NOAA and were provided by the Kanzelh\"{o}he Observatory, University of Graz, Austria.
%H.D.C., Y.N.S. and S.L.M. were supported by the National Key R\&D Program of China 2021YFA1600502, 2021YFA1600503 (2021YFA1600500), 2022YFF0503800, and 2022YFF0503003 (2022YFF0503000).
%H.D.C. and J.Z. acknowledge the NSFC support from grant numbers 11790304, 11790300 and 12073001, 12073042, 12350004, 12273061, and the Key Programs of the Chinese Academy of Sciences (QYZDJ- SSW-SLH050).
%S.L.M. was also supported by the NSFC (11790301 and 11941003).
%Y.N.S. also acknowledges the support of NSFC (11790302, 11790300 and 41761134088).
%G.P.Z. was also supported by the NSFC (11873059), the Key Research Program of Frontier Sciences, CAS (grant No. ZDBS-LY-SLH013), Beijing Natural Science Foundation (1202022), and Yunnan Academician Workstation of Wang Jingxiu (No. 202005AF150025). 
%Y.D.S. acknowledges the NSFC support (12173083, 11922307 and 11773068) and the Yunnan Science Foundation for Distinguished Young Scholars (202101AV070004).
\end{acknowledgments}

%\section{Author contributions}
%H.D.C. was responsible for the planning, coordination, image processing, analysis of the observations and writing of most of the manuscript. S.L.M., Y.N.S., G.P.Z., Y.D.S., C.X., W.N.T. and J.Z. contributed to the data analysis and discussion. In addition, S.L.M. plotted Figure 7 of the manuscript. Y.N.S. performed tentative reconstruction of the magnetic field of the prominence. G.P.Z. provided help to the writing of the manuscript. E.P. contributed to the data interpretation and discussion and improved the manuscript. L.F. provided suggestions about the figure, contributed to the discussion and improved the manuscript.

\begin{figure}[ht!]
\epsscale{1}
\plotone{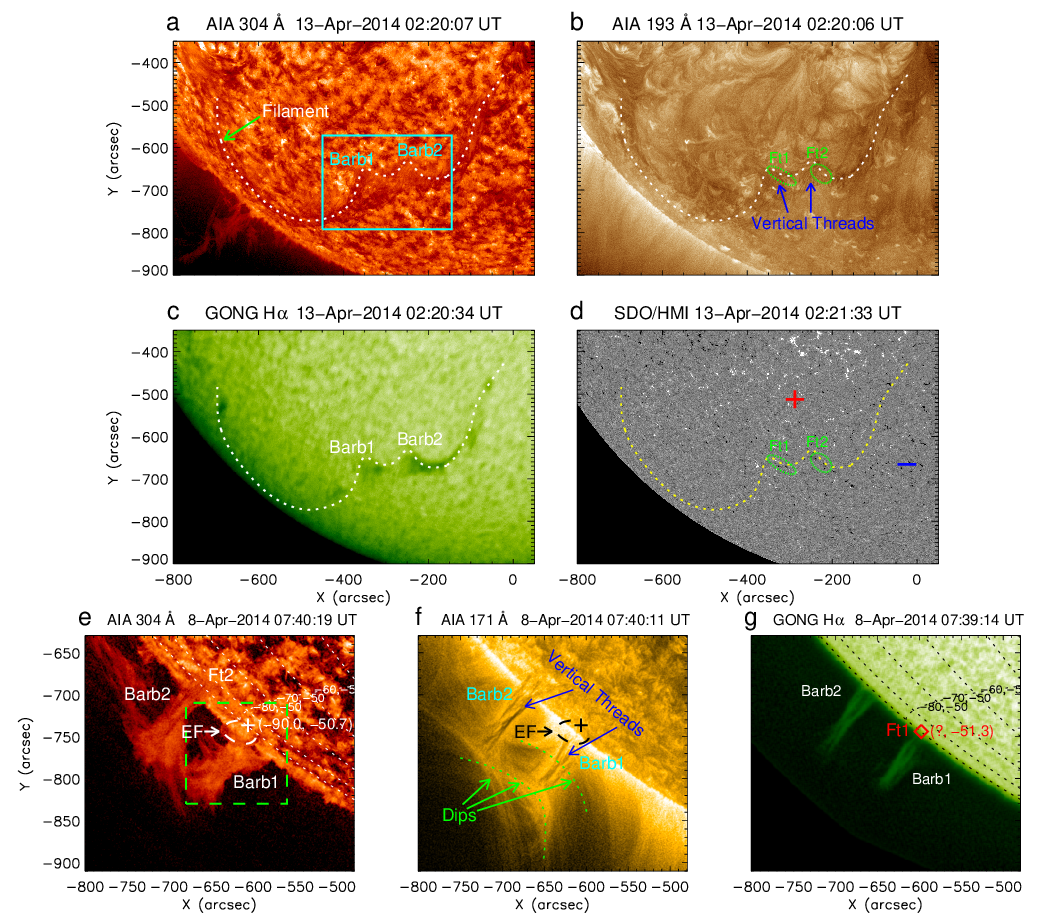}
\caption{
(a)--(c) SDO AIA 304 \AA, 193 \AA\ and GONG H${\alpha}$ intensity data on April 13 show that the filament had a spine structure with two apparent barbs (Barb1 and Barb2).
The box in panel (a) approximately indicates the field of view (FOV) of panels (e), (f), and (g).
(d) SDO HMI line-of-sight photospheric magnetogram with a field strength range of [-80, 80]~G.
The plus and minus sign in panel (d) represent the major positive and negative magnetic fields on the two sides of the filament, respectively.
The two ellipses in panels (b) and (d) indicate the locations of Barb1 and Barb2's footpoints (Ft1 and Ft2).
(e)--(g) The two barbs in the AIA 304 \AA, 171 \AA, and GONG H${\alpha}$ waveband appeared as two vertical feet of a prominence at the solar limb on April 8.
The dashed rectangle in panel (e) corresponds to the FOV of Fig. ~4 and 5.
The short dashed curves in panels (e) and (g) on the solar disk are the longitudinal and latitudinal lines.
The long dashed curves and plus signs in panels (e) and (f) denote the erupting filament EF and its source region, respectively. 
The green dashed curves in panel (f) outline some dip structures at the top of or across the barbs.
The blue arrows in panels (b) and (f) point to the vertical threads of the barbs.
The diamond sign in panel (g) indicates the footpoint of Barb1.
 An animation (Video~1) of the AIA 304 \AA\ (panels (a)), 193 \AA\ (panels (b)), 171 \AA, and  211 \AA\ (not shown in the static figure) images are available. 
 The animation proceeds from 07:00 UT on 2014 April 8 to 05:50 UT on 2014 April 13, illustrating the long-term evolutions of Barb1 and Barb2 in the filament.
 %``EF'' in panel (e) and (f) refer to a small-scale erupting filament. 
%The location of EF eruption source is denoted by the plus signs in panel (e) and (f).
 \label{fig:f1}}
\end{figure}

\begin{figure}[ht!]
\epsscale{0.9}
\plotone{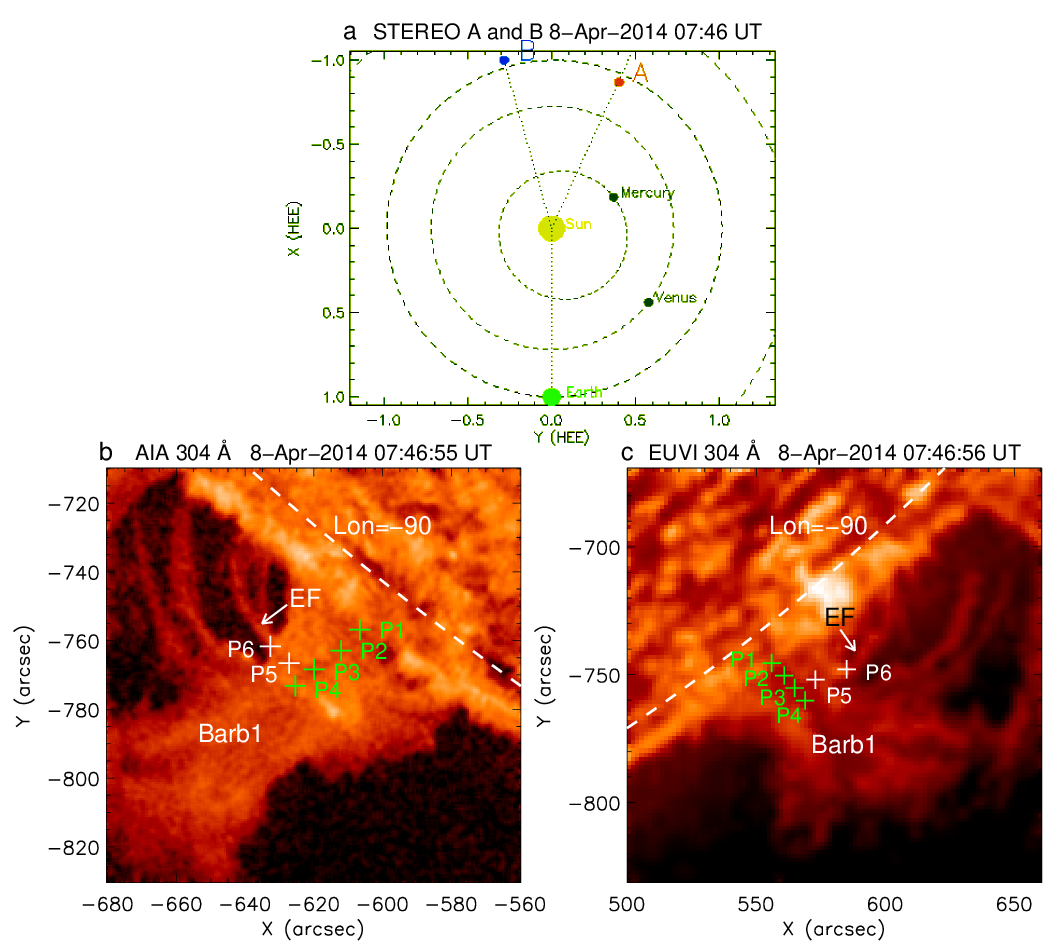}
\caption{
(a) The positions of STEREO A and B at 07:46 UT on 2014 April 8, obtained from the online STEREO orbit tool of the website (https://stereo-ssc.nascom.nasa.gov/where.shtml).
The SDO AIA (b) and STEREO B EUVI (c) 304 \AA\ intensity images show Barb1 and EF at the same time.
The dashed lines in panels (b) and (c) are the longitudinal lines of $-$90\degr.
The green (P1--P4) and white (P5--P6) plus signs denote the points at Barb1 and EF selected for the longitude calculations.
\label{fig:fig2}}
\end{figure}

\begin{figure}[ht!]
\epsscale{0.8}
\plotone{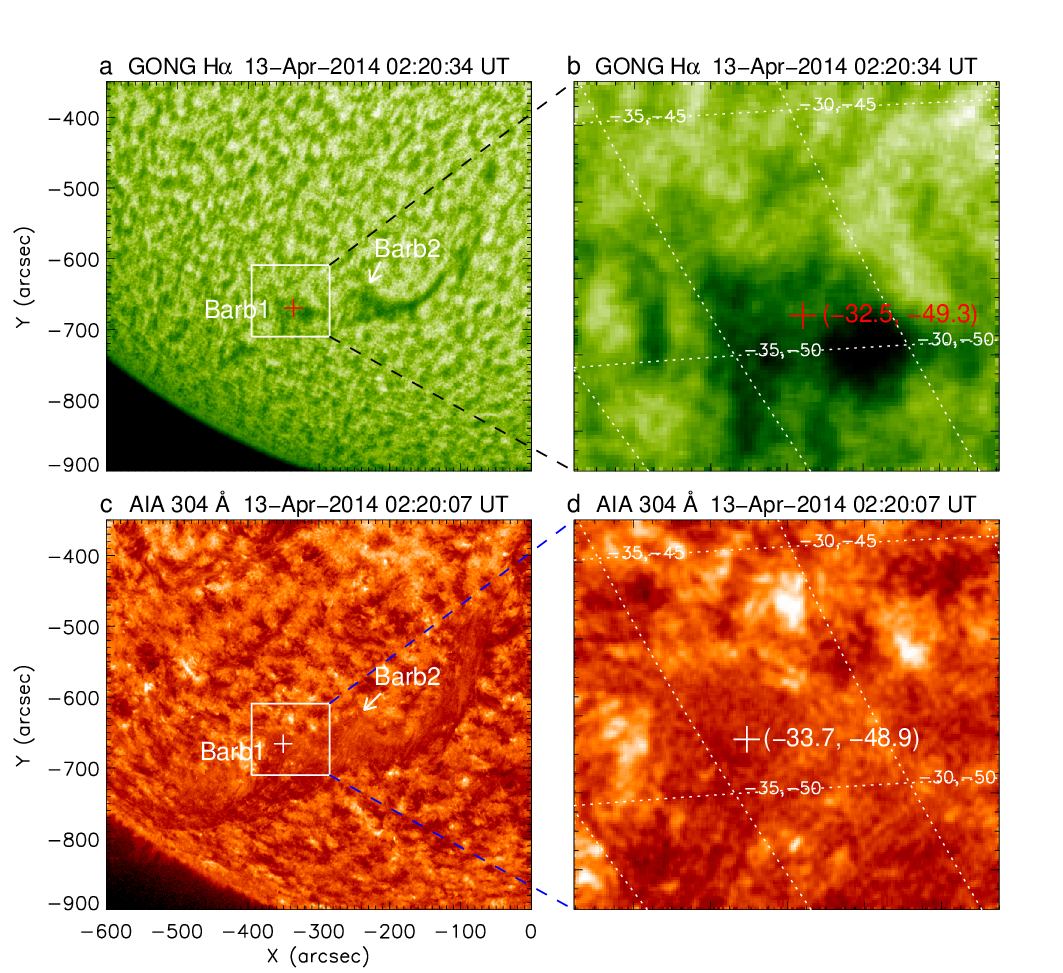}
\caption{
The GONG H${\alpha}$ (a--b) and SDO AIA 304 \AA\  (c--d) intensity images at 02:20 UT on April 13. 
The rectangles in panels (a) and (c) correspond to the FOVs of panels (b) and (d).
The plus signs approximately indicate the footpoint of Barb1.
\label{fig:fig3}}
\end{figure}

\begin{deluxetable*}{cccc} % {cclcDhh} 
\tablenum{1}
\tablecaption{Heliographic coordinates of Barb1 and the erupting filament EF\label{tab:tab1}}
\tablewidth{0pt}
\tablehead{
\colhead{Point} & \colhead{Longitude} & \colhead{Latitude} & \colhead{Radial Distance} \\
\colhead{} & \colhead{($\arcdeg$)} & \colhead{($\arcdeg$)} & \colhead{($R_\sun$)} 
%solar radii
%\nocolhead{Latitude}
%\multicolumn2c{(kpc)}
}
%\decimalcolnumbers
\startdata
\sidehead{Barb1:}
P1 & -91.44200 &  -51.43780  &   1.01516 \\
P2 & -91.81230  & -51.32870  &   1.02321 \\
P3 & -92.83280 &  -51.12050  &   1.03241 \\
P4 & -93.13420  & -51.01850  &   1.03992 \\
\sidehead{EF:}
P5 & -91.91580  & -50.74870  &   1.03505 \\
P6 & -89.80030  & -50.48840  &   1.03670  \\
\enddata
%\tablecomments{}
\end{deluxetable*}

\begin{figure}[ht!]
\epsscale{1.}
\plotone{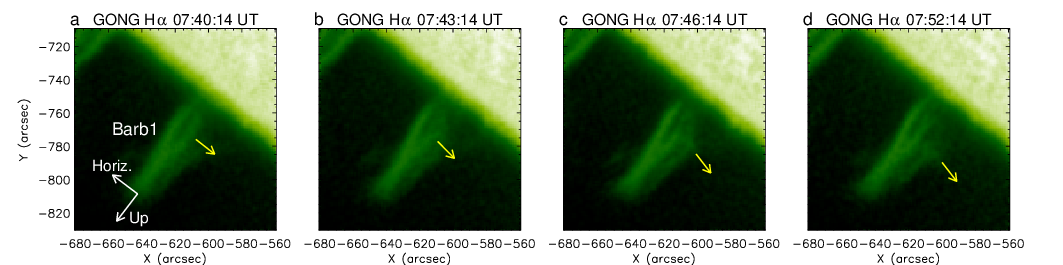}
\caption{
The GONG H${\alpha}$ line-center intensity data display the disturbance of Barb1 due to the filament EF's eruption on April 8. For more information, see Video~2.
The yellow arrows indicate the motion direction of the Barb1 material disturbed by EF.
The white arrows in panel (a) denote the local upward and horizontal direction.
\label{fig:fig4}}
\end{figure}

\begin{figure}[ht!]
\plotone{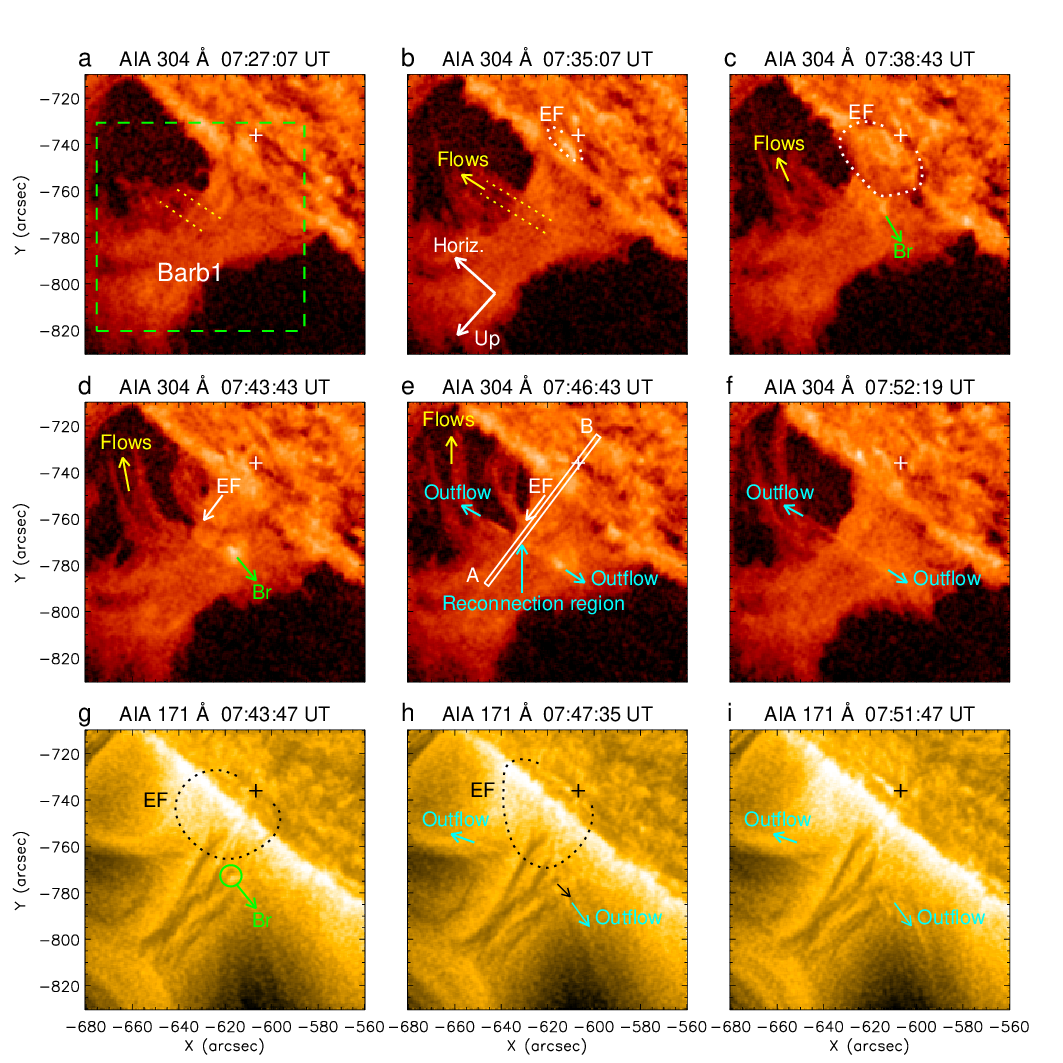}
\caption{
%SDO AIA observation of the filament EF eruption event on 2014 April 8.
%the interactions of the small erupting filament with Ar and Barb1.
The filament EF interacted with Barb1 in the AIA 304 \AA\ ((a)--(f)) and 171 \AA\ ((g)--(i)) channel.
The square in panel (a) shows the FOV of panels (a--c) of Fig.~6.
The plus signs in the diagram indicate the source region of the EF eruption. 
The yellow dotted lines in panels (a) and (b) indicate some horizontal structures in front of Barb1 and its motion (Flows) are denoted by the yellow arrows in panels (b)--(e).
The white (panels (b) and (c)) and black (panels (g) and (h)) dotted curves outline the EF.
``Br'' marks an EUV brightening and its motion is indicated by the green arrows in panels (c), (d), and (g).
The narrow box in panel (e) is the slit ``A--B'' along which the AIA 304 \AA\ time-distance diagram was plotted in panel (a) of Fig.~7.
The cyan arrows in panels (e), (f), (h), and (i) represent the motion directions of the reconnection outflows.
The black arrow in panel (h) indicate the disturbance and motion of some absorption features in Barb1.
The white arrows in panel (b) denote the local upward and horizontal direction.
 An animation (Video~3) of the AIA 304 \AA\ (panels (a)--(f)), 171 \AA\ (panels (g)--(i)), and 211 \AA\ (not shown in the static figure) images are available. The animation proceeds from 07:20 UTC to 08:20 UTC on 2014 April 8, illustrating the small filament EF's eruption and its interaction with Barb1.
\label{fig:fig5}}
\end{figure}

\begin{figure}[ht!]
\plotone{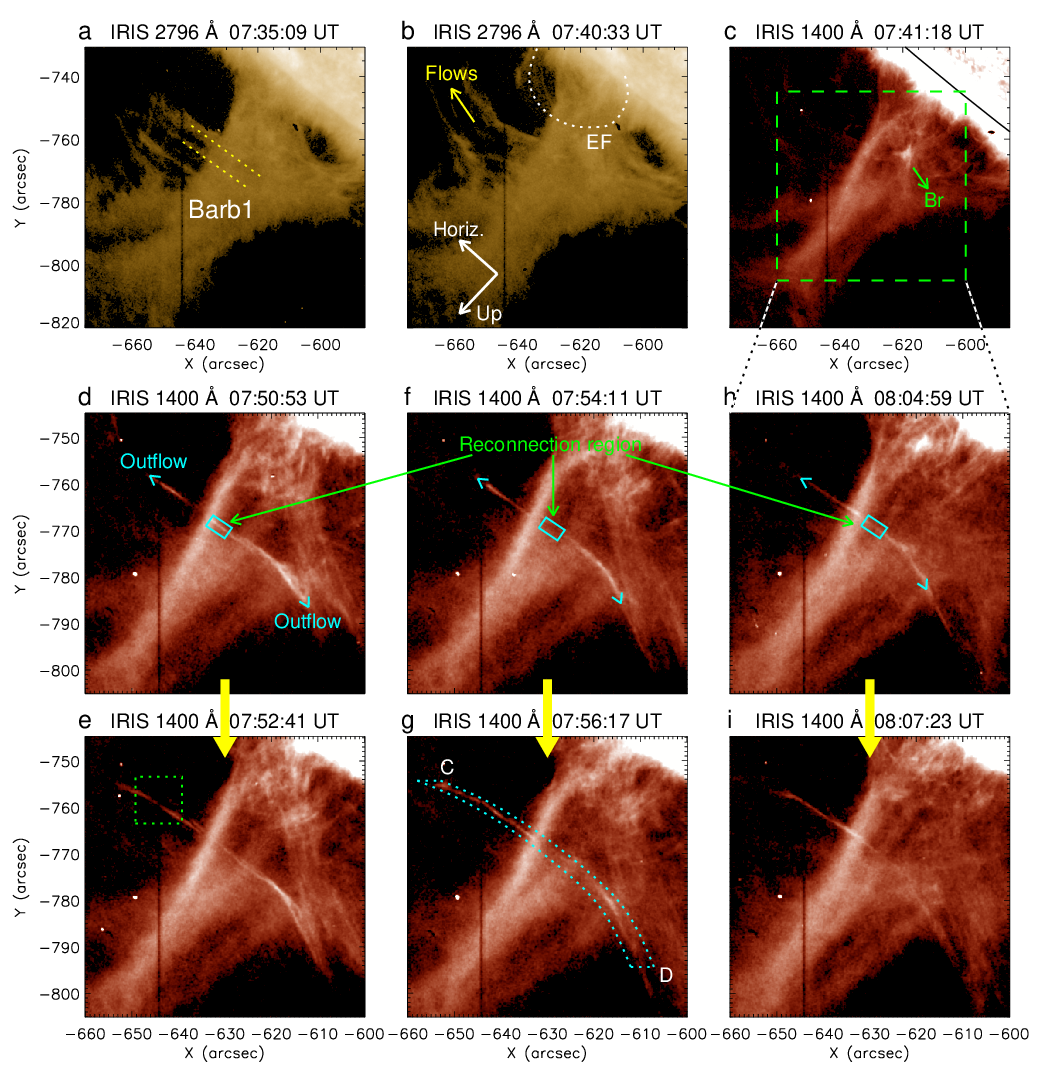}
\caption{
%IRIS SJI observation of the filament EF eruption.
(a--c) The early evolution of the EF eruption in the IRIS 2796 \AA\ and 1400 \AA\ channel.
The yellow dotted lines in panels (a) indicate the horizontal structures in front of Barb1 and its motion (Flows) are denoted by the yellow arrow in panels (b).
The white dotted curve in panel (b) outline part of EF due to the limited FOV of the IRIS SJI observations.
``Br'' marks a UV brightening and its motion is indicated by the green arrow in panels (c).
The square in panel (c) corresponds to the FOV of panels (d)--(i).
(d--i) Three pairs of bidirectional outflows due to the intermittent reconnections between the fields of filament EF and magnetic dip in Barb1.
The cyan arrows in panels (d), (f), and (h) indicate the motion directions of the reconnection outflows.
%The oblique rectangles in \textbf{d}, \textbf{f} and \textbf{h} roughly mark the central spots of the second reconnection.
The square in panel (e) is the FOV of panels (d) and (g) in Fig.~8.
The curved cut in panel (g) indicates the slit ``C--D'' along which the 1400 \AA\ time-distance map was plotted in the panel (d) of Fig.~7.
The white arrows in panel (b) denote the local upward and horizontal direction.
For more details, see Video~4.
\label{fig:fig6}}
\end{figure}

\begin{figure}[ht!]
\plotone{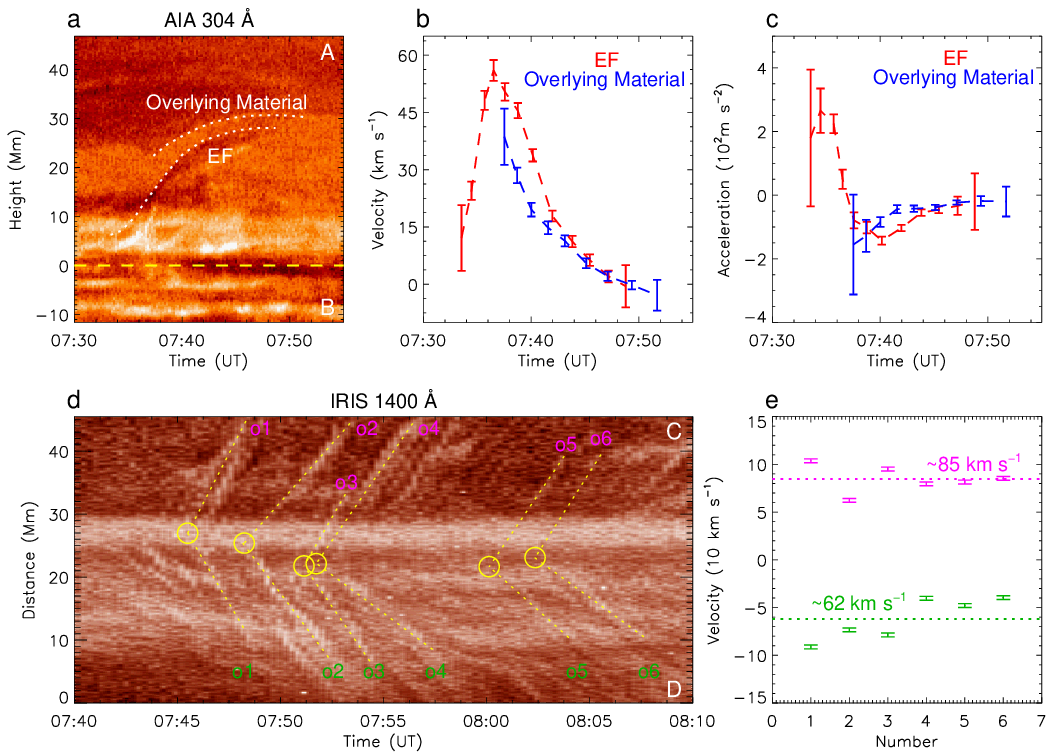}
\caption{
Dynamics of the erupting filament EF in AIA 304 \AA\ and the IRIS SJI bidirectional reconnection outflows.
(a) The AIA 304 \AA\ time-distance map shows the height variations of filament EF and the overlying compressed prominence material in Barb1 during the eruption.
The yellow dashed horizontal line indicates the solar surface.
The velocity-time (panel (b)) and acceleration-time (panel (c)) profiles of EF (red) and the overlying prominence material (blue) with 1$\sigma$ (standard deviation) error bars. 
%(c) The time variations of the accelerations of EF (red) and the compressed prominence material (blue) with 1$\sigma$ uncertainty.
%acceleration-time profiles of EF and the compressed prominence material.
%\textbf{c}, Similar to \textbf{b}, but for the acceleration-time profiles.
(d) IRIS 1400 \AA\ time-distance slit image enhanced by the MGN method presents six pairs of bidirectional reconnection outflows (o1--o6). 
%caused by the reconnection between EF and the magnetic dip in Barb1.
The circles mark the spots where the reconnection outflows originated.
(e) The projected velocity distribution of the reconnection outflows with 1$\sigma$ error bars.
The purple and green dotted lines represent the mean projected velocities of the reconnection outflows moving toward C and D, respectively.
\label{fig:fig7}}
\end{figure}

\begin{figure}[ht!]
\epsscale{0.7}
\plotone{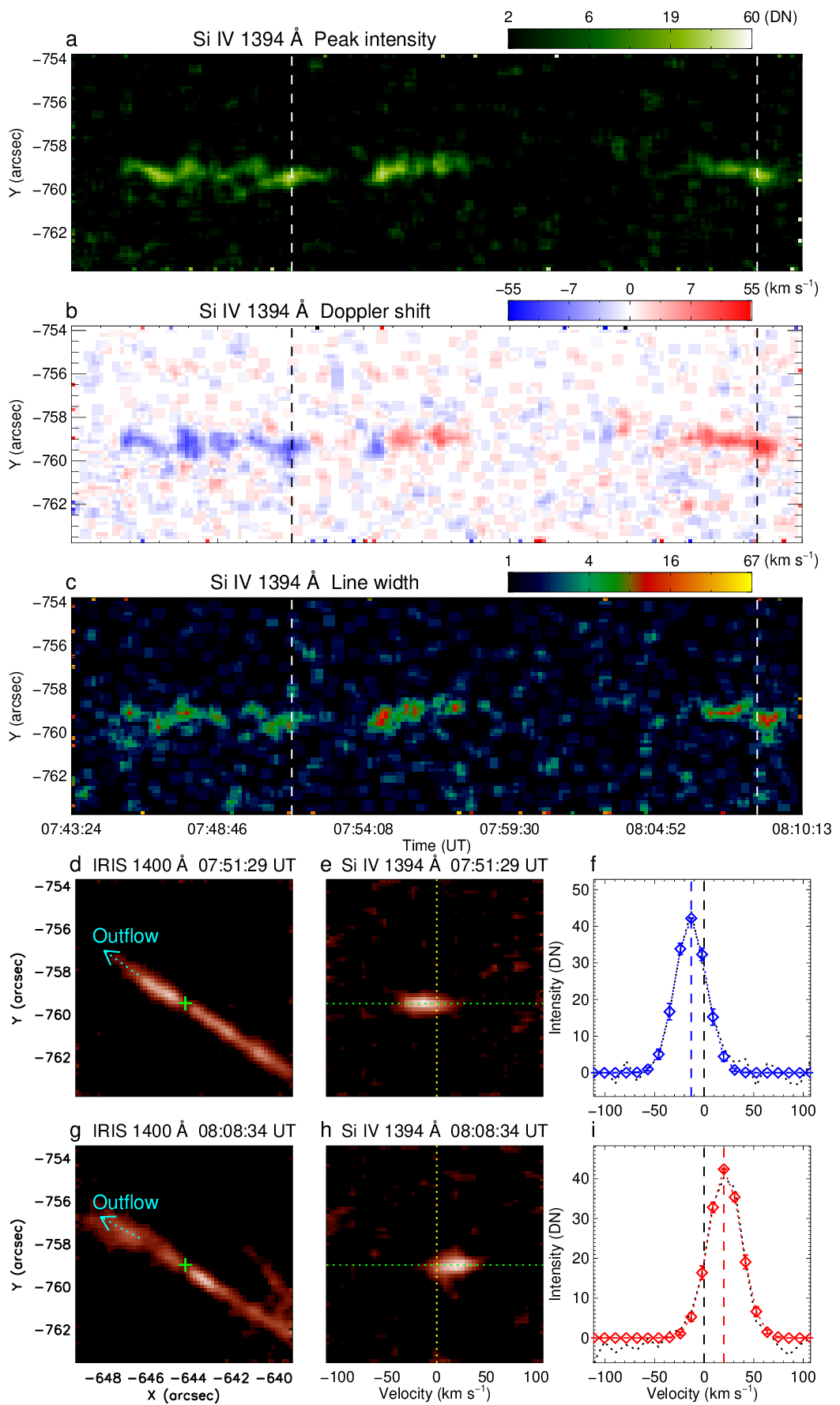}
\caption{
IRIS spectral observations of the reconnection outflow.
(a--c) Temporal evolutions of peak intensity, Doppler shift and line width derived from a single-Gaussian fitting to the spectra of \ion{Si}{4} 1393.76 \AA\ line at T = 0.08 MK.
%\textbf{a}, Temporal evolution of peak intensity derived from a single-Gaussian fitting to the spectra of Si IV 1393.78 \AA\ line.
%\textbf{b--c}, Similar to \textbf{a}, but for Doppler shift and line width, respectively.
The two vertical lines in panels (a), (b) and (c) indicate the time when the two reconnection outflows in panels (d) and (g) were observed by IRIS.
%the IRIS 1400 \AA\ intensity images in \textbf{d} and \textbf{g} were taken, respectivley.
The IRIS 1400 \AA\ imaging data, \ion{Si}{4} 1394 \AA\ line spectrum and profile of the reconnection outflow at 07:51:29 UT (panels (d)--(f)) and 08:08:35 UT (panels (g)--(i)), respectively. 
The cyan arrows in panels (d) and (g) indicate the motion direction of the two reconnection outflows.
%(g--i), The IRIS 1400 \AA\ imaging data, \ion{Si}{4} 1394 \AA\ line spectrum and profile of the reconnection outflow at 08:08:35 UT, respectively. 
%\textbf{d}, The IRIS 1400 \AA\ intensity data at 07:51:29 UT shows one outflow crossed the IRIS spectrometer slit.
%\textbf{e--f}, The corresponding spectra and profile of the Si IV 1393.78 \AA\ line.
%\textbf{g--i}, Similar to \textbf{d--f}, but for the IRIS data at 08:08:35 UT.
The plus signs in panels (d) and (g) mark the sites where the reconnection outflows crossed the spectrometer slit.
%intersections of the outflows and the IRIS slit.
The blue (red) dotted curve in panels (f) and (i) is the single-Gaussian fitting profile with 1$\sigma$ uncertainty.
\label{fig:fig8}}
\end{figure}

\begin{figure}[ht!]
\plotone{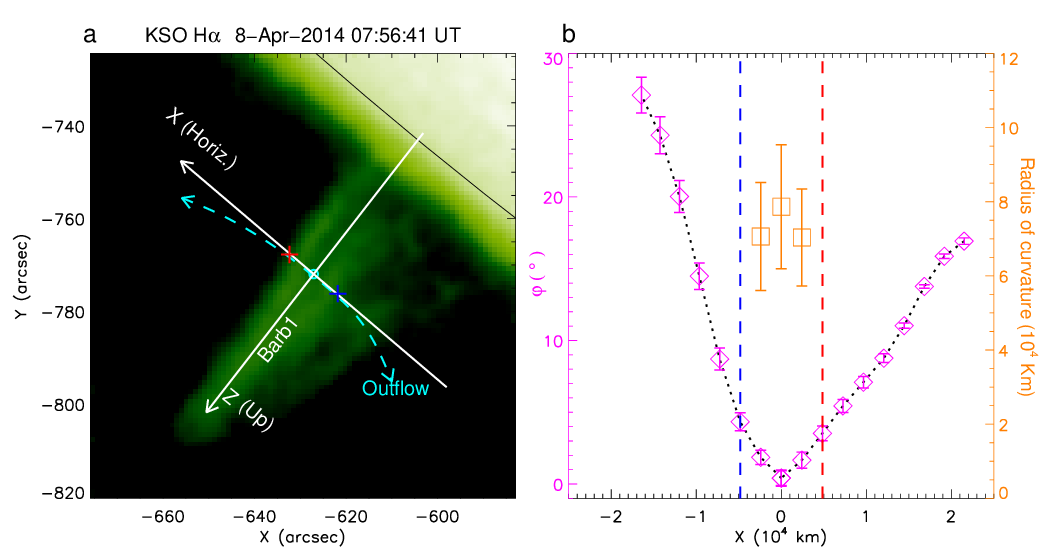}
\caption{
The field orientation of the magnetic dip indicated by reconnection outflows across Barb1.
(a) The cyan dashed profile of the bidirectional reconnection outflows at 07:56:17 UT (see Fig6.~(g)) overlaid on the KSO H${\alpha}$ intensity data.
$X$- and $Z$- axes represent the local horizontal and upward directions, respectively.
The black curve denotes the solar limb.
(b) The variations of the angle ($\varphi$, diamond and dotted curve) between the dip field and x-axis and the curvature radius (square) projected in the plane of the sky of the dip bottom with 1$\sigma$ error bars.
The plus signs in panel (a) and vertical lines in panel (b) indicate the boundaries of Barb1.
\label{fig:fig9}}
\end{figure}
%\textcolor{red}{}

\begin{figure}[ht!]
\plotone{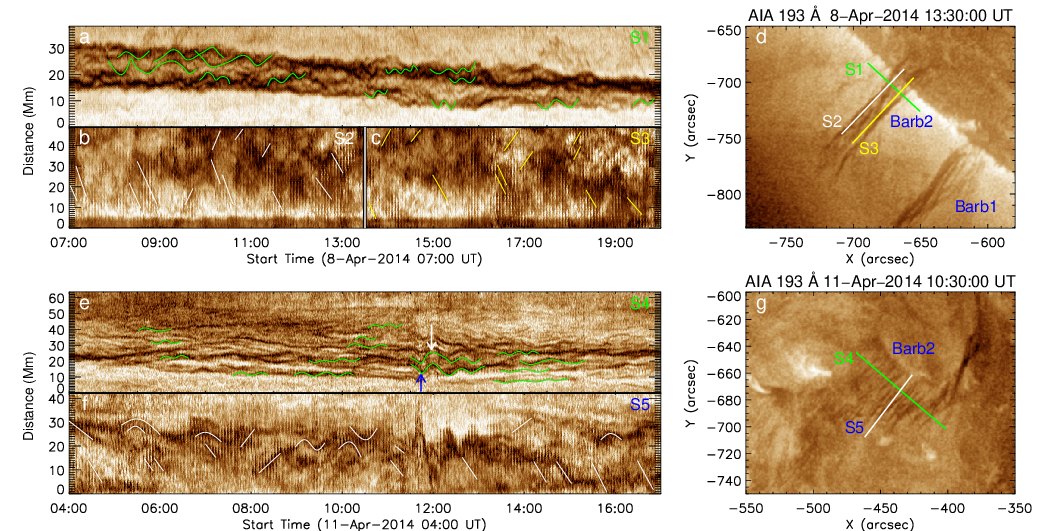}
\caption{
(a) The AIA 193 \AA\ time-distance map along the slit ``S1'' shows the horizontal oscillations (green curves) in Barb2 from 07:00 to 20:00 UT on April 8, 2014.
(b)--(c) The AIA 193 \AA\ time-distance map along the slits ``S2'' and ``S3'' exhibits the vertical flows in Barb2 from 07:00 to 13:30 UT (white lines) and from 13:30 to 20:00 UT (yellow lines) on April 8, respectively.
(d) The AIA 193 \AA\ image displays the positions of the slits S1, S2, and S3 on Barb2.
(e)--(g) Similar to (a)--(d), but for the slits ``S4'' and ``S5'' on Barb2 from 04:00 to 17:00 UT on April 11.
The white curves in panel (f) indicate the continuous processes of the up- and down-flows. 
The white and blue arrows in panel (e) denote the stronger oscillations caused by a coronal wave originating from a distant eruption.
Fore more information, see Video~5.
\label{fig:fig10}}
\end{figure}

\begin{deluxetable*}{cccccc clccc} % {cclcDhh} 
\tablenum{2}
\tablecaption{Kinetics of Horizontal Oscillations and Vertical Flows in Barb2\label{tab:tab2}}
\tablewidth{0pt}
\tablehead{
\colhead{Date} &\vline& \colhead{Horizontal} & \colhead{Amplitude} & \colhead{Period}
& \colhead{Velocity} & \vline& \colhead{Vertical       } & \colhead{Distance} & \colhead{Duration} & \colhead{Velocity} \\
\colhead{} &\vline& \colhead{Oscillation} & \colhead{$(Mm)$} & \colhead{$(minutes)$}
& \colhead{$(km~s^{-1})$} & \vline & \colhead{Flow} & \colhead{$(Mm)$} & \colhead{$(minutes)$} & \colhead{$(km~s^{-1})$} 
%solar radii
%\nocolhead{Latitude}
%\multicolumn2c{(kpc)}
%\cline{3-5}
%\vline
}
%\decimalcolnumbers
\startdata
%\sidehead{Barb1:}
08-Apr-2014 &\vline& &1.4 &  29  & 3.5 &\vline&  Up-flow & 7.2 & 12 & 10 \\
 (At solar limb)&\vline& & &    &  &\vline&  Down-flow & 11.1 & 15 & 13 \\
 \cline{1-11}
11-Apr-2014 &\vline& &0.9 &  30  & 2.3 &\vline&  Up-flow & 4.3 & 16 & 4 \\
 (On solar disk)&\vline& & &    &  &\vline&  Down-flow & 6.7 & 18 & 6 \\
%\cutinhead{M-class flares}
\enddata
\tablecomments{The oscillations' amplitudes and velocities and the vertical flows' distances and velocities are the mean projection values.}
\end{deluxetable*}

\begin{figure}[ht!]
\plotone{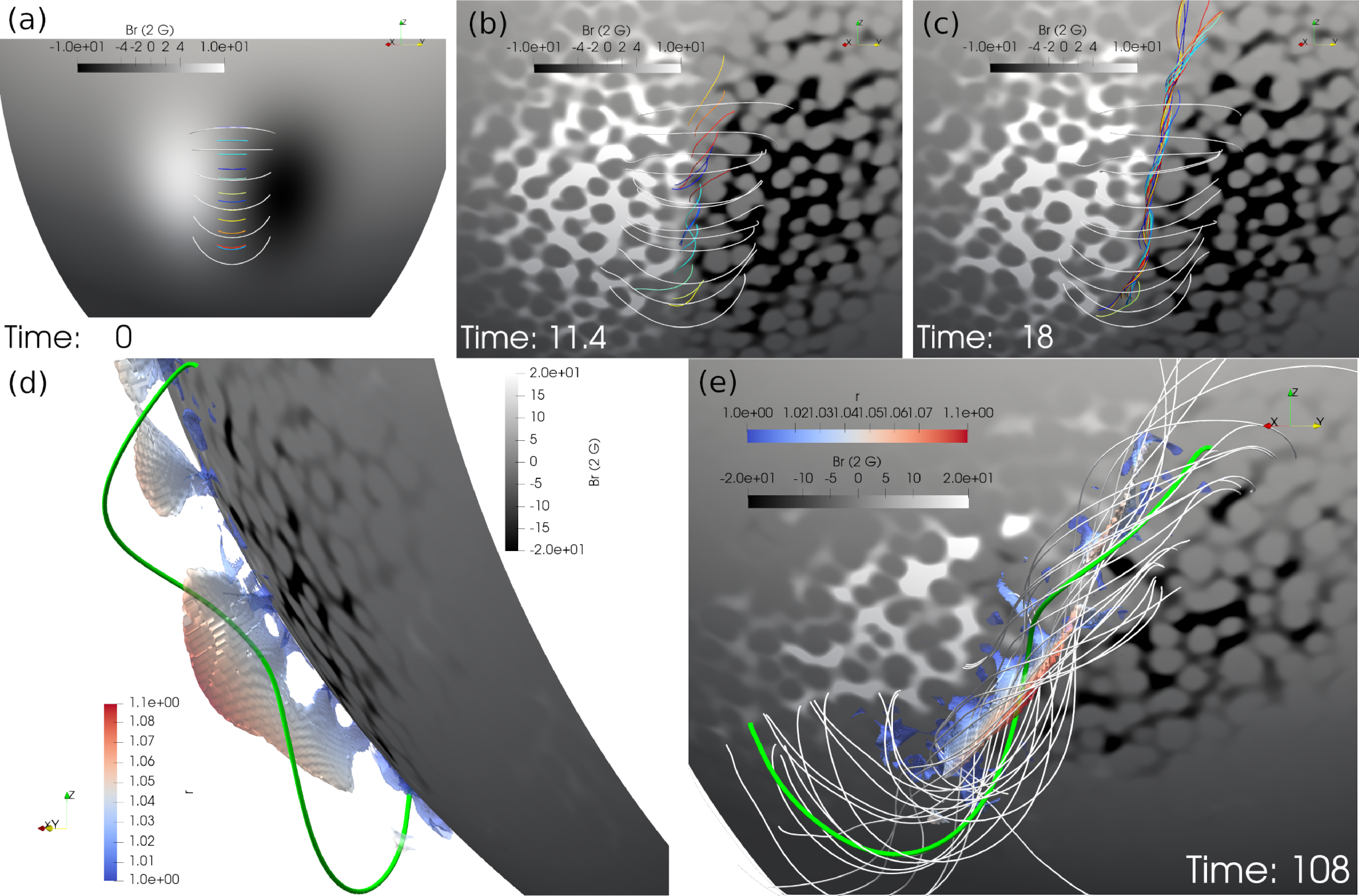}
\caption{Magnetofrictional simulation of quiescent prominence magnetic field driven by supergranulations. (a) Magnetic field lines of 
a bipolar potential magnetic field at time 0 as the initial condition in the spherical coordinates of a partial southern hemisphere. 
(b) Sheared magnetic arcades in rainbow colors and overlying potential magnetic loops in white at time 11.4. (c) Thin magnetic flux rope 
consisting of helical magnetic field lines in rainbow colors and overlying potential magnetic loops in white at time 18. (d) Side view of 
the simulated prominence at the solar limb presented by translucent magnetic dip regions colored by solar radii in red-to-blue colors 
and a green helical magnetic field line passing through the middle region of a prominence foot. (e) Top view of the same data as in (d) 
as a filament on the solar disk with additional white helical field lines presenting the structure of a magnetic flux rope at time 108. The 
bottom surfaces of all panels in black and white presenting photospheric magnetograms $B_r$. 
\label{fig:fig11}}
\end{figure}

\begin{figure}[ht!]
\plotone{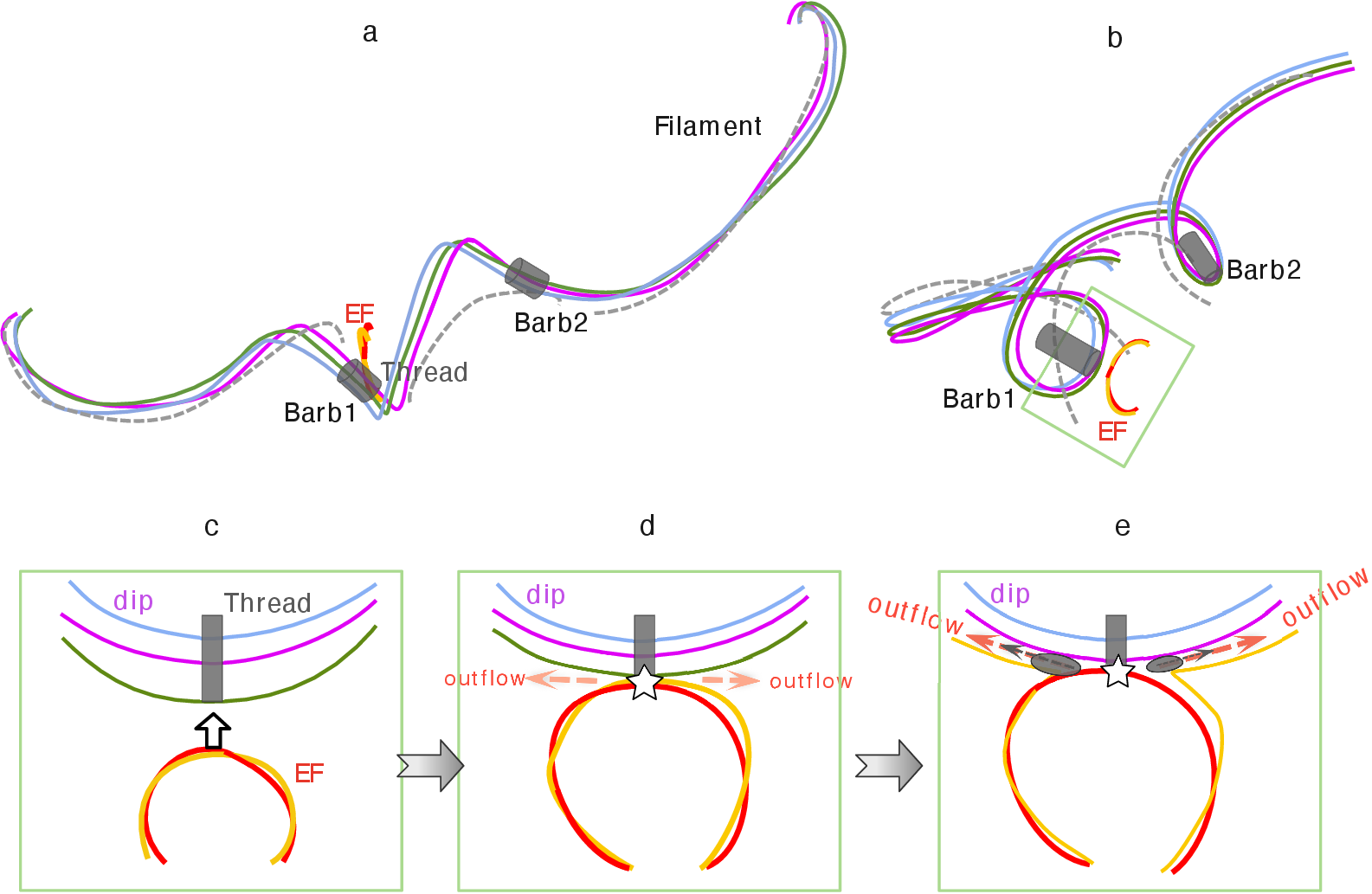}
\caption{
Schematic diagram of magnetic field structure of the larger-scale filament and its reconnection with the fields of the filament EF under Barb1.
(a--b) Top- and side-view of the filament field structure, respectively.
%\textbf{b}, Side-view of the filament field structure.
(c--e) The filament EF's eruption and its interactions with the magnetic dip in Barb1.
%interaction of EF with the magnetic arcade (Ar) and the magnetic dip in Barb1. 
The box in panel (b) marks the FOV of  panels (c--e).
The dark grey cylinders represent the vertical threads in the barbs.
\label{fig:fig12}}
\end{figure}

\end{CJK*}

\begin{thebibliography}{}
\bibitem[Anzer \& Heinzel(2007)]{anzer07} Anzer, U. \& Heinzel, P.\ 2007, \aap, 467, 1285
\bibitem[Aulanier \& Demoulin(1998)]{aulanier98a} Aulanier, G. \& Demoulin, P.\ 1998, \aap, 329, 1125
\bibitem[Aulanier et al.(1998)]{aulanier98b} Aulanier, G., Demoulin, P., van Driel-Gesztelyi, L., et al.\ 1998, \aap, 335, 309
\bibitem[Barczynski et al.(2021)]{barczynski21} Barczynski, K., Schmieder, B., Peat, A.~W., et al.\ 2021, \aap, 653, A94

\bibitem[Berger et al.(2012)]{berger12} Berger, T.~E., Liu, W., \& Low, B.~C.\ 2012, \apjl, 758, 2, L37

\bibitem[Berger et al.(2008)]{berger08} Berger, T.~E., Shine, R.~A., Slater, G.~L., et al.\ 2008, \apjl, 676, L89
\bibitem[Berger et al.(2011)]{berger11} Berger, T., Testa, P., Hillier, A., et al.\ 2011, \nat, 472, 197
\bibitem[Bi et al.(2014)]{bi14} Bi, Y., Jiang, Y., Yang, J., et al.\ 2014, \apj, 790, 100
\bibitem[Bi et al.(2020)]{bi20} Bi, Y., Yang, B., Li, T., et al.\ 2020, \apjl, 891, L40
\bibitem[Bommier \& Leroy(1998)]{bommier98} Bommier, V. \& Leroy, J.~L.\ 1998, IAU Colloq. 167: New Perspectives on Solar Prominences, 150, 434
\bibitem[Bommier et al.(1986)]{bommier86} Bommier, V., Sahal-Brechot, S., \& Leroy, J.~L.\ 1986, \aap, 156, 79
\bibitem[Chae(2010)]{chae10} Chae, J.\ 2010, \apj, 714, 618
\bibitem[Chae et al.(2005)]{chae05} Chae, J., Moon, Y.-J., \& Park, Y.-D.\ 2005, \apj, 626, 574
\bibitem[Chen et al.(2014)]{chenh14} Chen, H., Zhang, J., Cheng, X., et al.\ 2014, \apjl, 797, L15
\bibitem[Chen et al.(2020a)]{chenh20} Chen, H., Zhang, J., De Pontieu, B., et al.\ 2020a, \apj, 899, 19
\bibitem[Chen et al.(2016)]{chenh16} Chen, H., Zhang, J., Li, L., et al.\ 2016, \apjl, 818, L27
\bibitem[Chen et al.(2015)]{chenh15} Chen, H., Zhang, J., Ma, S., et al.\ 2015, \apjl, 808, L24
\bibitem[Chen et al.(2020b)]{chenp20} Chen, P.-F., Xu, A.-A., \& Ding, M.-D.\ 2020b, Research in Astronomy and Astrophysics, 20, 166
\bibitem[Chen et al.(2024)]{cheny24} Chen, Y., Mandal, S., Peter, H., et al.\ 2024, \aap, 692, A119
\bibitem[De Pontieu et al.(2014)]{depontieu14} De Pontieu, B., Title, A.~M., Lemen, J.~R., et al.\ 2014, \solphys, 289, 2733
\bibitem[Dud{\'\i}k et al.(2012)]{dudik12} Dud{\'\i}k, J., Aulanier, G., Schmieder, B., et al.\ 2012, \apj, 761, 9
\bibitem[{{Fisher} {et~al.}(2020){Fisher}, {Kazachenko}, {Welsch}, {Sun},
  {Lumme}, {Bercik}, {DeRosa}, \& {Cheung}}]{Fisher2020}
{Fisher}, G.~H., {Kazachenko}, M.~D., {Welsch}, B.~T., {et~al.} 2020, \apjs, 248, 2
\bibitem[Freeland \& Handy(1998)]{freeland98} Freeland, S.~L. \& Handy, B.~N.\ 1998, \solphys, 182, 497
\bibitem[Gibson(2018)]{gibson18} Gibson, S.~E.\ 2018, Living Reviews in Solar Physics, 15, 7
\bibitem[Gun{\'a}r et al.(2018)]{gunar18} Gun{\'a}r, S., Dud{\'\i}k, J., Aulanier, G., et al.\ 2018, \apj, 867, 115
\bibitem[Guo et al.(2021)]{guo21} Guo, Y., Hou, Y., Li, T., et al.\ 2021, \apjl, 911, L9
\bibitem[Haerendel \& Berger(2011)]{haerendel11} Haerendel, G. \& Berger, T.\ 2011, \apj, 731, 82
\bibitem[Harvey et al.(1996)]{harvey96} Harvey, J.~W., Hill, F., Hubbard, R.~P., et al.\ 1996, Science, 272, 1284
\bibitem[Hillier et al.(2012a)]{hillier12a} Hillier, A., Berger, T., Isobe, H., et al.\ 2012a, \apj, 746, 120
\bibitem[Hillier et al.(2012b)]{hillier12b} Hillier, A., Isobe, H., Shibata, K., et al.\ 2012b, \apj, 756, 110
\bibitem[Hou et al.(2024)]{houz24} Hou, Z., Tian, H., Madjarska, M.~S., et al.\ 2024, \aap, 687, A190
\bibitem[Howard \& Harvey(1970)]{howard70} Howard, R. \& Harvey, J.\ 1970, \solphys, 12, 23
\bibitem[Jing et al.(2003)]{jing03} Jing, J., Lee, J., Spirock, T.~J., et al.\ 2003, \apjl, 584, 2, L103
\bibitem[Kaiser et al.(2008)]{kaiser08} Kaiser, M.~L., Kucera, T.~A., Davila, J.~M., et al.\ 2008, \ssr, 136, 5
\bibitem[{{Keppens} {et~al.}(2023){Keppens}, {Popescu Braileanu}, {Zhou},
  {Ruan}, {Xia}, {Guo}, {Claes}, \& {Bacchini}}]{Keppens2023}
{Keppens}, R., {Popescu Braileanu}, B., {Zhou}, Y., {et~al.} 2023, \aap, 673, A66
\bibitem[Kippenhahn \& Schl{\"u}ter(1957)]{kippenhahn57} Kippenhahn, R. \& Schl{\"u}ter, A.\ 1957, \zap, 43, 36
\bibitem[Labrosse et al.(2010)]{labrosse10} Labrosse, N., Heinzel, P., Vial, J.-C., et al.\ 2010, \ssr, 151, 243
\bibitem[Lemen et al.(2012)]{lemen12} Lemen, J.~R., Title, A.~M., Akin, D.~J., et al.\ 2012, \solphys, 275, 17
\bibitem[Levens et al.(2016)]{levens16} Levens, P.~J., Schmieder, B., L{\'o}pez Ariste, A., et al.\ 2016, \apj, 826, 164
\bibitem[Li et al.(2018)]{lid18} Li, D., Shen, Y., Ning, Z., et al.\ 2018, \apj, 863, 192
\bibitem[Li \& Zhang(2013)]{lil13} Li, L. \& Zhang, J.\ 2013, \solphys, 282, 147
\bibitem[Li \& Zhang(2012)]{lit12} Li, T. \& Zhang, J.\ 2012, \apjl, 760, L10
\bibitem[Lin et al.(2009)]{lin09} Lin, Y., Soler, R., Engvold, O., et al.\ 2009, \apj, 704, 870
\bibitem[Liu \& Ofman(2014)]{liuw14} Liu, W. \& Ofman, L.\ 2014, \solphys, 289, 3233
\bibitem[{Liu \& Xia(2022)}]{Liu2022}
Liu, Q., \& Xia, C.\ 2022, \apjl, 934, L9
\bibitem[L{\'o}pez Ariste et al.(2006)]{lopez06} L{\'o}pez Ariste, A., Aulanier, G., Schmieder, B., et al.\ 2006, \aap, 456, 725
\bibitem[Low et al.(2012a)]{low12a} Low, B.~C., Berger, T., Casini, R., et al.\ 2012a, \apj, 755, 34
\bibitem[Low et al.(2012b)]{low12b} Low, B.~C., Liu, W., Berger, T., et al.\ 2012b, \apj, 757, 21

\bibitem[Luna et al.(2012)]{luna12b} Luna, M., D{\'\i}az, A.~J., \& Karpen, J.\ 2012, \apj, 757, 1, 98
\bibitem[Luna et al.(2024)]{luna24} Luna, M., Joshi, R., Schmieder, B., et al.\ 2024, \aap, 691, A354
\bibitem[Luna \& Karpen(2012)]{luna12a} Luna, M. \& Karpen, J.\ 2012, \apjl, 750, 1, L1
\bibitem[Luna et al.(2017)]{luna17} Luna, M., Su, Y., Schmieder, B., et al.\ 2017, \apj, 850, 143
\bibitem[Luna et al.(2016)]{luna16} Luna, M., Terradas, J., Khomenko, E., et al.\ 2016, \apj, 817, 2, 157

\bibitem[Mackay et al.(2010)]{mackay10} Mackay, D.~H., Karpen, J.~T., Ballester, J.~L., et al.\ 2010, \ssr, 151, 333
\bibitem[Mackay et al.(2020)]{mackay20} Mackay, D.~H., Schmieder, B., L{\'o}pez Ariste, A., et al.\ 2020, \aap, 637, A3
\bibitem[Martens \& Zwaan(2001)]{martens01} Martens, P.~C. \& Zwaan, C.\ 2001, \apj, 558, 872
\bibitem[Martin(1998)]{martin98} Martin, S.~F.\ 1998, \solphys, 182, 107
%\bibitem[Martin et al.(2009)]{martin09} Martin, S.~F., Panasenco, O., Agah, Y., et al.\ 2009, The Second Hinode Science Meeting: Beyond Discovery-Toward Understanding, 415, 183
\bibitem[Mart{\'\i}nez Gonz{\'a}lez et al.(2015)]{martinez15} Mart{\'\i}nez Gonz{\'a}lez, M.~J., Manso Sainz, R., Asensio Ramos, A., et al.\ 2015, \apj, 802, 3
%12. Mart\'{i}nez Gonz\'{a}lez, M. J., Asensio Ramos, A., Arregui, I., Collados, M., Beck, C. \& de la Cruz Rodr\'{i}guez, J. On the magnetism and dynamics of prominence legs hosting tornadoes. Astrophys. J. 825, 119 (2016).
\bibitem[Morgan \& Druckm{\"u}ller(2014)]{morgan14} Morgan, H. \& Druckm{\"u}ller, M.\ 2014, \solphys, 289, 2945
\bibitem[Ouyang et al.(2020)]{ouyang20} Ouyang, Y., Chen, P.~F., Fan, S.~Q., et al.\ 2020, \apj, 894, 64
\bibitem[Panesar et al.(2013)]{panesar13} Panesar, N.~K., Innes, D.~E., Tiwari, S.~K., et al.\ 2013, \aap, 549, A105
\bibitem[Parenti(2014)]{parenti14} Parenti, S.\ 2014, Living Reviews in Solar Physics, 11, 1
\bibitem[Pesnell et al.(2012)]{pesnell12} Pesnell, W.~D., Thompson, B.~J., \& Chamberlin, P.~C.\ 2012, \solphys, 275, 3
\bibitem[Petrie \& Low(2005)]{petrie05} Petrie, G.~J.~D. \& Low, B.~C.\ 2005, \apjs, 159, 288
\bibitem[Priest(2014)]{priest14} Priest, E. (ed.)\ 2014, Magnetohydrodynamics of the Sun (Cambridge: Cambridge Univ. Press)
%40. Kippenhahn, R. \& Schl\"{u}ter, A. Eine theorie der solaren filamente. Zeitschrift f\"{u}r Astrophysik Bd. 43, 36--62 (1957).
\bibitem[Rees-Crockford et al.(2024)]{rees24} Rees-Crockford, T., Scullion, E., Khomenko, E., et al.\ 2024, \apj, 974, 64

\bibitem[R{\'e}gnier \& Amari(2004)]{regnier04} R{\'e}gnier, S. \& Amari, T.\ 2004, \aap, 425, 345
\bibitem[R{\'e}gnier et al.(2011)]{regnier11} R{\'e}gnier, S., Walsh, R.~W., \& Alexander, C.~E.\ 2011, \aap, 533, L1

\bibitem[Scherrer et al.(2012)]{scherrer12} Scherrer, P.~H., Schou, J., Bush, R.~I., et al.\ 2012, \solphys, 275, 207
\bibitem[Schmieder et al.(2010)]{schmieder10} Schmieder, B., Chandra, R., Berlicki, A., et al.\ 2010, \aap, 514, A68
\bibitem[Schmieder et al.(2015)]{schmieder15} Schmieder, B., L{\'o}pez Ariste, A., Levens, P., et al.\ 2015, Polarimetry, 305, 275
\bibitem[Schmieder et al.(2017)]{schmieder17} Schmieder, B., Zapi{\'o}r, M., L{\'o}pez Ariste, A., et al.\ 2017, \aap, 606, A30
\bibitem[Schou et al.(2012)]{schou12} Schou, J., Scherrer, P.~H., Bush, R.~I., et al.\ 2012, \solphys, 275, 229
\bibitem[Shen et al.(2014)]{shen14} Shen, Y., Liu, Y.~D., Chen, P.~F., et al.\ 2014, \apj, 795, 130.
\bibitem[Shen et al.(2015)]{shen15} Shen, Y., Liu, Y., Liu, Y.~D., et al.\ 2015, \apjl, 814, L17
\bibitem[Shen et al.(2019)]{shen19} Shen, Y., Qu, Z., Yuan, D., et al.\ 2019, \apj, 883, 104
\bibitem[Shen et al.(2022)]{shen22} Shen, Y., Zhou, X., Duan, Y., et al.\ 2022, \solphys, 297, 20
%37. Peter, H. et al. Hot explosions in the cool atmosphere of the Sun. Science 346, 1255726 (2014).
\bibitem[Su \& van Ballegooijen(2012)]{suyn12} Su, Y. \& van Ballegooijen, A.\ 2012, \apj, 757, 168.
\bibitem[Su et al.(2015)]{suyn15} Su, Y., van Ballegooijen, A., McCauley, P., et al.\ 2015, \apj, 807, 144
\bibitem[Su et al.(2012)]{suy12} Su, Y., Wang, T., Veronig, A., et al.\ 2012, \apjl, 756, L41
\bibitem[Tandberg-Hanssen(1995)]{tandberg95} Tandberg-Hanssen, E.\ 1995, in The Nature of Solar Prominences, ed. I. Appenzeller et al. (Astrophysics and Space Science Library, Vol. 199; Dordrecht: Kluwer)
Astrophysics and Space Science Library, vol. 199, Dordrecht: Kluwer Academic Publishers, |c1995
\bibitem[Terradas et al.(2013)]{terradas13} Terradas, J., Soler, R., D{\'\i}az, A.~J., et al.\ 2013, \apj, 778, 1, 49
\bibitem[Tian et al.(2014)]{tian14} Tian, H., DeLuca, E.~E., Cranmer, S.~R., et al.\ 2014, Science, 346, 1255711
\bibitem[Tian et al.(2015)]{tian15} Tian, H., Young, P.~R., Reeves, K.~K., et al.\ 2015, \apj, 811, 139
\bibitem[van Ballegooijen(2004)]{van04} van Ballegooijen, A.~A.\ 2004, \apj, 612, 519
\bibitem[van Ballegooijen \& Martens(1989)]{van89} van Ballegooijen, A.~A. \& Martens, P.~C.~H.\ 1989, \apj, 343, 971
\bibitem[Vial \& Engvold(2015)]{vial15} Vial, J.-C. \& Engvold, O.\ 2015, Solar Prominences, 415
\bibitem[Wang et al.(2016)]{wang16} Wang, B., Chen, Y., Fu, J., et al.\ 2016, \apjl, 827, 2, L33
\bibitem[Wuelser et al.(2004)]{wuelser04} Wuelser, J.-P., Lemen, J.~R., Tarbell, T.~D., et al.\ 2004, \procspie, 5171, 111
\bibitem[Xia \& Keppens(2016)]{xia16} Xia, C. \& Keppens, R.\ 2016, \apjl, 825, L29
\bibitem[Xia et al.(2014a)]{xia14a} Xia, C., Keppens, R., Antolin, P., et al.\ 2014a, \apjl, 792, L38
\bibitem[Xia et al.(2014b)]{xia14b} Xia, C., Keppens, R., \& Guo, Y.\ 2014b, \apj, 780, 130
\bibitem[{{Xia} {et~al.}(2018){Xia}, {Teunissen}, {El Mellah}, {Chan{\'e}}, \&
  {Keppens}}]{Xia2018}
{Xia}, C., {Teunissen}, J., {El Mellah}, I., {Chan{\'e}}, E., \& {Keppens}, R.
  2018, \apjs, 234, 30
%6. Awasthi, A. K., Liu, R. \& Wang, Y. M. Double-decker filament configuration revealed by mass motions. Astrophys. J. 872, 109 (2019).
\bibitem[Yan et al.(2015)]{yan15} Yan, X.~L., Xue, Z.~K., Pan, G.~M., et al.\ 2015, \apjs, 219, 17
\bibitem[Yan et al.(2025)]{yan25} Yan, X., Xue, Z., Wang, J., et al.\ 2025, \apj, 981, 2, 139

%50. Xue, Z. K., Yan, X. L., Yang, L. H., Wang, J. C. \& Zhao, L. Observing formation of flux rope by tether-cutting reconnection in the sun. Astrophys. J. Lett. 840, L23 (2017).
\bibitem[Yang et al.(2024a)]{yangb24} Yang, B., Yang, J., Bi, Y., et al.\ 2024a, \apjl, 970, L3
\bibitem[Yang et al.(2024b)]{yangj24} Yang, J., Chen, H., Hong, J., et al.\ 2024b, \apj, 964, 7
\bibitem[Yang et al.(2024c)]{yangl24} Yang, L., Yan, X., Xue, Z., et al.\ 2024c, \mnras, 528, 1094
\bibitem[Yokoyama \& Shibata(1995)]{yokoyama95} Yokoyama, T. \& Shibata, K.\ 1995, \nat, 375, 42

\bibitem[Zhang et al.(2012)]{zhangq12} Zhang, Q.~M., Chen, P.~F., Xia, C., et al.\ 2012, \aap, 542, A52
\bibitem[Zhang et al.(2013)]{zhangq13} Zhang, Q.~M., Chen, P.~F., Xia, C., et al.\ 2013, \aap, 554, A124
\bibitem[Zhang \& Ji(2018)]{zhangq18} Zhang, Q.~M. \& Ji, H.~S.\ 2018, \apj, 860, 113
\bibitem[Zhang et al.(2017)]{zhangq17} Zhang, Q.~M., Li, D., \& Ning, Z.~J.\ 2017, \apj, 851, 1, 47

\bibitem[Zheng(2024)]{zheng24} Zheng, R.\ 2024, Proceedings of the Royal Society of London Series A, 480, 20230950
\bibitem[Zheng et al.(2018)]{zheng18} Zheng, R., Chen, Y., Huang, Z., et al.\ 2018, \apj, 861, 108
\bibitem[Zhou et al.(2021)]{zhou21} Zhou, C., Xia, C., \& Shen, Y.\ 2021, \aap, 647, A112
\bibitem[Zhou et al.(2018)]{zhou18} Zhou, Y.-H., Xia, C., Keppens, R., et al.\ 2018, \apj, 856, 2, 179

\bibitem[Zirker et al.(1998)]{zirker98} Zirker, J.~B., Engvold, O., \& Martin, S.~F.\ 1998, \nat, 396, 440
%;------references in Methods--------
%Kiepenheuer, K. O. 1953, in The Sun, ed. G. P. Kuiper (Chicago, IL: Univ. Chicago Press), 322
\end{thebibliography}
\end{document}